\documentclass[useAMS,usenatbib]{mnras}
\usepackage{amsmath}
\usepackage{appendix}
\usepackage{graphicx}
\usepackage{xfrac}
\usepackage{caption}
\usepackage{array}
\usepackage{longtable}
\usepackage{pdflscape}
\usepackage[figuresright]{rotating}
\usepackage{booktabs}
\usepackage[usenames, dvipsnames]{color}
\usepackage{subcaption}
\usepackage[flushleft]{threeparttable}
\usepackage[export]{adjustbox}
\usepackage{hyperref}
\allowdisplaybreaks

\usepackage{float}

\usepackage{media9}
\usepackage{animate}
\usepackage{etoolbox}
\usepackage{makecell}

\emergencystretch=\maxdimen
\title[ProGeny II]{ProGeny II: the impact of libraries and model configurations on inferred galaxy properties in SED fitting }
\author[Bellstedt et al.]
{Sabine Bellstedt$^{1}$\thanks{Email: sabine.bellstedt@uwa.edu.au} \& Aaron S. G. Robotham$^{1}$
\\
$^{1}$ ICRAR, The University of Western Australia, 7 Fairway, Crawley WA 6009, Australia\\
}

\begin{document}

\date{}

\pagerange{\pageref{firstpage}--\pageref{lastpage}} \pubyear{2024}

\maketitle

\label{firstpage}

\begin{abstract}
	
We use a volume-complete sample of $\sim$8,000 galaxies from the GAMA survey to characterise the impact of stellar population libraries (SPLs) and model configurations on the resulting inferred galaxy properties from Spectral Energy Distribution (SED) fitting. 
We compare a fiducial SPL from \textsc{ProGeny} (a new tool that can generate SPLs quickly and flexibly) against five other commonly used SPLs using the SED-fitting code \textsc{ProSpect}. 
The impact of selecting each SPL is compared to the consequence of changing the model implementation in the SED fitting process, including the implementation of metallicity evolution versus a fixed or constant metallicity, and a functional parametric star formation history (SFH) versus a stepwise parametric (or ``non-parametric") SFH. 
Furthermore, we use \textsc{ProGeny} to assess the impact of sub-SPL choices, including isochrone selection, stellar spectra selection, and IMF selection. 
Through a comparison of derived stellar masses, star formation rates, metallicities, ages, and the inferred cosmic star formation history (CSFH), we rank the impact of varying choices. Overall the assumption of a solar metallicity creates the greatest biases, with a substantial impact also caused by the choice of a specific SPL. 
To recover a CSFH most consistent with observations, we advocate for the use of the fiducial implementation with a skewed Normal functional form for the SFH, and an evolving metallicity, although we note that all studied SPLs underestimate the peak in the CSFH. 

\end{abstract}


\section{Introduction}

After data collection, the first step to any observational study of galaxies is a measurement of their properties.
There is much that can be measured, such as the brightness at specific wavelengths, angular size, and redshift. 
With a redshift and an assumed cosmology, observed properties can be transformed to intrinsic ones, such as the galaxy luminosity and physical size. 
Much of modern galaxy evolution moves well past these more basic measurements, and relies strongly on the inference of the more fundamental properties of galaxies, such as the stellar mass, star formation rate, and age of a galaxy's stellar population. 
Such properties cannot be extracted through a simple \textit{measurement}, and instead relies on decades of research to provide modelling capability to rather \textit{infer} these properties. 

Stellar population sysnthesis (SPS) modelling uses knowledge about the evolutionary track of stars (isochrones), combined with a stellar initial mass function \citep[IMF, for example][]{salpeter1955, kroupa2002, chabrier2003} and a set of stellar spectra at every evolutionary stage of a star to build up a stellar population generatively \citep[for example, ][]{tinsley1976, tinsley1978, bruzual1993}. 

SPS models are capable of producing what are known as Simple Stellar Populations (SSPs), which for a given age and metallicity provide a template for the stellar population. A library of such SSPs should cover the entire stellar population, and is referred to as a Stellar Population Library (SPL). 
These SPLs \citep[of which many flavours exist, including for example][]{leitherer1999, bruzual2003, maraston2005, vazdekis2016} are used as a foundational input to model the observed Spectral Energy Distributions (SEDs) of galaxies via their broadband photometry, in a process named SED fitting. 
In SED fitting, a stellar population is described in terms of a star formation history (SFH) to account for its age distribution, and a corresponding metallicity history (ZH). This description is used to recreate the stellar emission of the galaxy by building up the appropriate SSPs. An additional dust component is typically modelled, which facilitates the attenuation of the stellar light, as well as the re-emission of this light in the far-infrared through an energy balance process.  
Much has been written on the topic of SED fitting, and some excellent reviews have been provided by \citet{walcher2011, conroy2013}. 
Today, many different SED fitting codes exist in the literature, including Magphys \citep{dacunha2008}, BayeSED \citep{han2012}, CIGALE \citep{boquien2019}, Prospector \citep{leja2017}, BAGPIPES \citep{carnall2018a}, BEAGLE \citep{chevallard2016}, and \textsc{ProSpect} \citep{robotham2020}. 

An alternative method to extract physical properties of galaxies is full spectral fitting, which is a fitting of SPLs to spectra rather than broadband photometry. Examples of such techniques include \textsc{pPXF} \citep{cappellari2004}, \textsc{Starlight} \citep{cidfernandes2005a}, and \textsc{Firefly} \citep[][which pursues an entirely non-parametric approach to spectral fitting]{wilkinson2017}. Despite having similar goals, the difference in wavelength coverage, required model resolution and computational demands are sufficiently large that spectral fitting and SED fitting are generally treated as distinct techniques, with related but separate challenges.   

Given the ubiquity of SED fitting techniques in deriving galaxy properties, and the importance of producing accurate properties, there has been an enormous effort devoted by the community to characterising the impacts of different assumptions in the SED fitting process. 
In recent times, a major focus of this effort has been on the manner in which the star formation history (SFH) within a model is parametrised, whether this be in a functional parametric form (typically referred to in the literature as ``parametric''), or a stepwise parametric form (typically referred to in the literature as ``non-parametric''\footnote{Note that we will not be adopting this nomenclature in this paper, as any parametrisation defined by free parameters in separate time bins is also parametric, making the term ``non-parametric'' confusing.}. ). Some SED fitting codes use exclusively smooth SFH functions (such as MagPhys \citealt{dacunha2008}, although a discontinuous burst can also be fitted), or stepwise SFHs (such as \textsc{Prospector} \citep{leja2017}), however increasingly codes are being designed with the functionality to use either (such as \textsc{ProSpect} \citealt{robotham2020} or \textsc{Bagpipes} \citealt{carnall2018a}).

Comparisons of the galaxy properties to have been derived by different SED fitting codes have been made in the past \citep[such as the extensive analysis by][]{pacifici2023}, and furthermore many studies have analysed the impact of individual elements of SED fitting. 
This includes \citet{pforr2012} who presented an early study assessing the impact of assumptions in SED fitting using mocks, \citet{carnall2019, leja2019a, lower2020, suess2022} who studied the impact of SFH parametrisations on stellar masses using Prospector and BAGPIPES, \citet{thorne2022} who studied the impact of the inclusion of AGN as a separate component, \citet{haskell2024} who studied the impact of star formation burstiness, and \citet{jones2022} who studied the impact of different dust models. 

A typical manner of characterising the impacts of SED fitting modelling assumptions (such as the form of the SFH parametrisation or the complexity of metallicity enrichment) is to generate a set of mock SEDs with known SFHs and metallicities using SPLs, and then extracting these known parameters through SED fitting processes \citep[see for example such approaches by][]{lower2020, suess2022}. 
While this is the only true way of quantifying the biases introduced by SED modelling assumptions, the use of a single SPL in generating the mocks, and then fitting them, means that the impact of uncertainties in the SPL itself cannot be gauged by such an analysis.

In this work, we aim to characterise the impact of the selected SPL on the final galaxy properties from SED fitting. 
We compare the SPL impact to that of other modelling choices, including the SFH parametrisation, and the metallicity modelling implementation. 
Finally, by generating individual SPLs with typical stellar spectra\footnote{In the context of \textsc{ProGeny} ``stellar spectra" encapsulates both empirical and theoretical spectra of stellar photospheres.}, isochrones, and IMFs, we further characterise the impact of these sub-SPL choices on the final galaxy properties derived using SED fitting. 
This sub-SPL comparison is facilitated through the development of a new SPL generation software \textsc{ProGeny}, which is outlined in a companion paper by \citet{robotham2024a}.

The outline of this paper is described as follows: Section \ref{sec:Data} outlines the data and stellar population libraries studied, and Section \ref{sec:Method} discusses the SED fitting methods applied in this work. Our results benchmarking the relative impacts of these stellar population libraries and commonly changed SED fitting configurations are presented in Section \ref{sec:Results}, and a demonstration of the impact of individual ingredients in the SPLs is presented in Section \ref{sec:OtherImpacts}. Finally we discuss the relative impacts of different assumptions in Section \ref{sec:Discussion}, and provide our main conclusions in Section \ref{sec:Conclusions}. 
 The cosmology assumed throughout this paper is $H_0 = 67.8\,\rm{km}\,\rm{s}^{-1}\,\rm{Mpc}^{-1}$,  $\Omega_m = 0.308$ and $\Omega_{\Lambda} = 0.692$ \citep[consistent with a Planck 15 cosmology][]{planckcollaboration2016}. 

\section{Data}
\label{sec:Data}

\subsection{GAMA}

We use the SED fitting code \textsc{ProSpect} \citep{robotham2020} on a volume-limited sample of galaxies from the GAMA survey \citep{driver2011, liske2015, driver2022} to assess the systematic differences in galaxy population recovery that result from the choice of stellar population library (SPL). 
The GAMA sample selected in this work is the $z<0.06$, $m_r < 19.65$ sample in the four main GAMA fields (G09, G12, G15, and G23) for which a spectroscopic redshift is available with a quality flag $NQ > 3$, corresponding to the selection within which the sample is 95\% complete \citep{driver2022}. This sample consists of 7,862 galaxies in total.  
Photometry is available in 20 bands from the FUV to the FIR \citep[as presented by][]{bellstedt2020a}, as taken from the \texttt{gkvInputCatv02} and  \texttt{gkvFIRv01} Data Management Units (DMU), and the redshifts and selection are taken from the \texttt{gkvScienceCatv02} DMU\footnote{Released publicly as part of GAMA DR4 \url{https://www.gama-survey.org/dr4/schema/}. }. 

\begin{figure*}
	\centering
	\includegraphics[width=180mm]{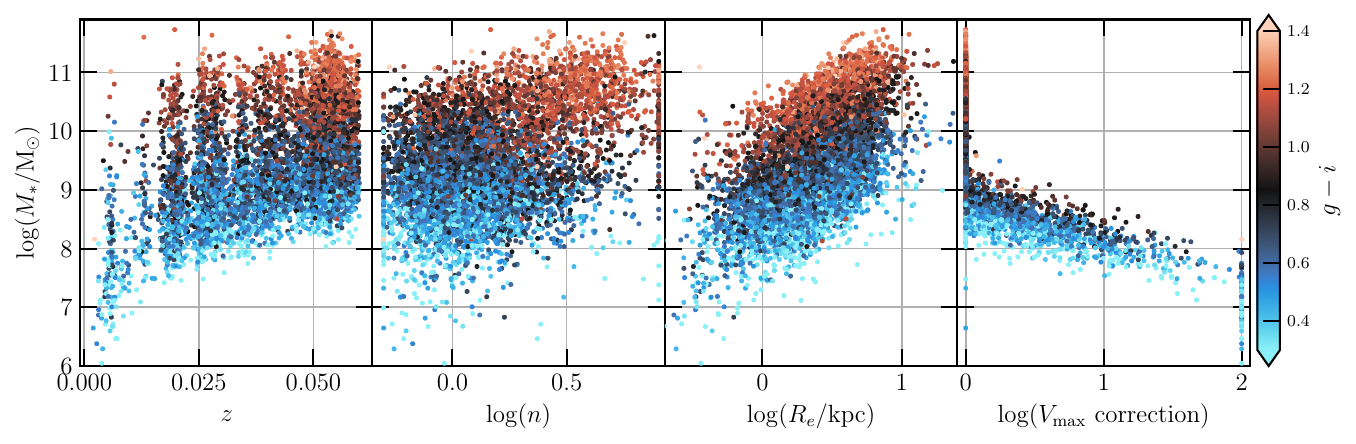}
	\caption{Properties of the GAMA sample analysed in this work, as derived using the single-sersic mode of \textsc{ProFuse} \citep{bellstedt2024}. We also present the Vmax corrections that are computed to account for incompleteness in the sample.}
	\label{fig:Sample_Properties}
\end{figure*}

An overview of the physical properties of the selected sample is provided in Fig. \ref{fig:Sample_Properties}. The stellar mass is plotted against the redshift, sersic index, and physical size. Stellar masses, sersic indices and sizes have all been derived by \citet{bellstedt2024} using \textsc{ProFuse}, where 2D light distributions have been simultaneously modelled across optical and near-infrared bands to derive the structural and SED-derived properties self-consistently. Note that the plotted stellar masses are independent from those measured in subsequent sections of this paper. 

Vmax values (which describe the maximum volume to which each galaxy is detectable within the sample selection) for each galaxy in the sample have been computed using \textsc{ProSpect} to identify the maximum redshift where the galaxy falls within the $m_r < 19.65$ limit. 
This has been done by directly shifting the redshift of the galaxy spectrum without considering any galaxy evolution (i.e., the assumed SFH remains the same). 
For galaxies that have a Vmax value less than the volume covered by the sample, a completeness correction can be defined as $V_{\rm survey}/V_{\rm max}$. The corrections computed for individual galaxies are presented in the fourth panel of Fig. \ref{fig:Sample_Properties}.

\subsection{Stellar Population Libraries}

\begin{table*}
	\centering
	\caption[SPLs]{Summary of stellar population libraries analysed in this work.  }
	\label{tab:SPLs}
	\begin{tabular}{@{}c | l l l c c }
		\hline
		SPL & Reference &  Isochrone  & Stellar Spectra & IMF &  Z Range  \\
		\hline
		\hline
		BC03 & \citet{bruzual2003}$^{*}$  &Padova 1994, Geneva & BaSeL, STELIB, Pickles & \citet{chabrier2003}&  [1e-4, 0.05]   \\
		\hline
		M05 & \citet{maraston2005}$^{**}$  & Frascati, Geneva 1992 & BaSeL, TP-AGB &  \citet{kroupa2002}&  [1e-3, 0.04]  \\
		\hline
		FSPS & \citet{conroy2009a}  &  Padova & MILES, BaSeL$^{\dagger}$  & \citet{chabrier2003}&  [2e-4, 0.03]  \\
		\hline
		BPASS & \citet{stanway2018}  & Cambridge STARS & BaSeL, C3K& \citet{chabrier2003}&  [1e-5, 0.04]   \\
		\hline
		CB19 & \citet{plat2019}$^{\ddagger}$  & PARSEC, COLIBRI &  MILES, BaSeL & \citet{chabrier2003}&  [1e-4, 0.06]   \\
		\hline
		ProGeny & \citet{robotham2024a}  & MIST & C3K  & \citet{chabrier2003} &  [2e-6, 0.063] \\
		\hline		
	\end{tabular}
	\begin{tablenotes}
		\small
		\item \textbf{Isochrones}: Padova 1994 \citep{girardi2002}, Geneva \citep{lejeune2001}, Frascati \citep{cassisi1997}, Geneva 1992 \citep{schaller1992},  Padova \citep{marigo2007, marigo2008}, Cambridge STARS \citep{eldridge2008}, PARSEC-COLIBRI \citep{marigo2017}, MIST \citep{dotter2016}. 
		\item \textbf{Stellar spectra}: BaSeL \citep{lejeune1997}, STELIB \citep{leborgne2003}, Pickles \citep{pickles1998}, MILES \citep{sanchez-blazquez2006}, C3K \citep{conroy2018}, TP-AGB \citep{lancon2002}. 
		\item \textbf{Comments:}
		\item $^{*}$ Note that the BC03hr version of BC03 was used in this work. 
		\item $^{**}$ The M05 version used includes red horizontal branch stars at all metallicities, resulting in no UV source at old ages. 
		\item $^{\dagger}$ Here the stellar spectra from MILES are used, but the wavelength extension into the UV and NIR is provided by BaSeL. 
		\item $^{\ddagger}$ CB19 provides an evolution of BC03, including updated spectra for hot massive stars that produce more He \textsc{II}-ionizing radiation, intended to better reproduce observations \citep{plat2019}.  
	\end{tablenotes}
\end{table*}

The six SPLs compared in this work are presented in Table \ref{tab:SPLs}, and include (in order of original release) BC03 \citep{bruzual2003}, M05 \citep{maraston2005}, FSPS \citep{conroy2009}, BPASS \citep{stanway2018}, CB19 (an update of \citealt{bruzual2003}, and introduced in \citealt{plat2019}), and \textsc{ProGeny}\footnote{Available publicly at \url{https://github.com/asgr/ProGeny}} \citet{robotham2024a}\footnote{While available for use within \textsc{ProGeny}, we have elected not to include the libraries EMILES \citep{vazdekis2016} and XSL \citep{verro2022} in this paper, due to their truncated wavelength range, and lack of availability of low-age templates. In general, these libraries are better suited for spectral analysis, rather than broadband SEDs. }. 
Each SPL has been constructed using stellar spectra and isochrones that originate from slightly different sources. 
Furthermore, the nature by which these libraries have been constructed (including interpolation schemes, and how different stellar spectra sets are prioritised) also vary. 
For further details on the exact construction of these SPLs, please see \citet{robotham2024a}. 

\section{Method}
\label{sec:Method}

We use the SED fitting code \textsc{ProSpect} \citep{robotham2020} to test and compare the relative impact of using each of the independent stellar population libraries (SPLs herein), the metallicity implementation (as this is a parameter that varies significantly in terms of its degeneracy with age with different SPLs), and broadly the SFH parametrisation. 
When fitting the photometric data from GAMA, we augment the measured uncertainties with the error floors presented in table 1 of \citet{bellstedt2020b}, to account for systematic photometric and modelling uncertainties.

\begin{table}
	\centering
	\caption[Configurations]{Summary of \textsc{ProSpect} configurations compared in this work for each SPL. The specific parameters fitted or fixed in each configuration are provided in Table \ref{tab:FreeParameters}.  }
	\label{tab:Configurations}
	\begin{tabular}{@{}c | c c }
		\hline
		Configuration & SFH & ZH \\
		\hline
		\hline
		Fiducial & snorm trunc & Evolving and free \\
		$Z_{\odot}$ & snorm trunc & Fixed Solar \\
		Constant Z & snorm trunc & Constant and free \\
		b5 SFH & b5 &  Evolving and free \\
		\hline		
	\end{tabular}
\end{table}

For each of the compared SPLs, we conduct SED fitting in four different modes, that are summarised in Table. \ref{tab:Configurations}. 
The fiducial configuration implemented here is consistent with that of \citet{bellstedt2020b, bellstedt2021}, in which a smooth functional form of the star formation history is selected (\texttt{massfunc\_snorm\_trunc}, a skewed Normal with a truncation to 0 at the early Universe, facilitating a range of potential SFHs, from rising, to constant, and declining, see \citet{robotham2020} for a demonstration). 
A metallicity evolution is also implemented, where the metal build-up is linearly mapped from the mass build-up, and the final metallicity is fitted as a free parameter (Zfinal). 

In the $Z_{\odot}$ configuration, the metallicity implementation is changed from the fiducial by fixing the metallicity at every epoch to be solar ($Z = 0.02$). While overly simplistic, this is still a popular simplification in many SED fitting applications \citep{yang2022, paspaliaris2023}. 

The constant Z configuration allows the metallicity to be fitted as a free parameter, however this value is taken as constant with time. This is more flexible than the $Z_{\odot}$ configuration, and is generally the standard approach for present-day SED fitting techniques \citep{leja2017, carnall2018a, iyer2019, johnson2021}. 

The final b5 SFH configuration employs the same metallicity evolution implementation as the fiducial model, however it employs a stepwise parametrisation for the star formation history (available in \textsc{ProSpect} as the \texttt{massfunc\_b5} function), more similar to the stepwise parametric SFHs used typically by codes like \textsc{Prospector} \citep{leja2017}. This parametriation divides the SFH into five time bins, where the star formation rate (SFR) is fitted independently in each bin. 
The \texttt{massfunc\_b5} parametrisation is identical to that outlined by \citet{robotham2020}. The five bins are divided by six lookback time boundaries relative to the observed epoch of the galaxy, at 0, 0.1, 1, 5, 9 in Gyr, and the upper boundary of the oldest bin is defined as 13.38 Gyr in absolute lookback time (which shifts relative to the SFH of any individual galaxy observed at different redshifts). These bin limits have been determined so that a constant shift in SFR would have a similar impact on the bolometric light output from each bin, meaning that a modification of SFR in each bin has a roughly similar impact on the resulting SED. 
The advantage of this over \texttt{massfunc\_snorn\_trunc} is that the SFH is not limited to being unimodal, and an old population can be fitted with the SFR rising again at recent times. 

Dust is attenuated in the SED according to the \citet{charlot2000} model, and re-emitted in the FIR according to a \citet{dale2014} dust emission model. 
Due to the low prevalence of AGN in the low-$z$ GAMA sample \citep[as demonstrated by][]{thorne2022}, we opt not to include AGN in the fit to limit the number of unnecessary free parameters. 

\begin{table}
	\centering
	\caption[FreeParameters]{Free and fixed parameters in each SED fitting configuration. Z$_{\rm min}$ refers to the minimum metallicity available from the SPL used. The Log column indicates whether the parameter was fitted in log space or not, and if no response is indicated then the parameter was set to be fixed (not free).  }
	\label{tab:FreeParameters}
	\begin{tabular}{@{}l | cc |  c}
	\hline
	Parameter & Log & Range/Value &Units\\
	\hline
	\hline
	\multicolumn{4}{l}{Fiducial configuration} \\[3pt]
	\hline
	mSFR  & Yes & [$-3$, 4] & $\rm M_{\odot}/{\rm yr}$ \\
	mpeak  & No & [$-2$, $13.4-t_{\rm lb}$]  & Gyr\\
	mperiod  & Yes & [$-1$, 1] & Gyr  \\
	mskew  & No & [$-1$, 1] & -- \\
	Zfinal  & Yes & [max($-4$, log(Z$_{\rm min}$)), $-1.3$]&  -- \\
	Zstart  & - & max($-4$, log(Z$_{\rm min}$))&  -- \\
	\hline
	\multicolumn{4}{l}{Constant Z configuration} \\[3pt]
	\hline
	mSFR  & Yes & [$-3$, 4] & $\rm M_{\odot}/{\rm yr}$ \\
	mpeak  & No & [$-2$, $13.4-t_{\rm lb}$]  & Gyr\\
	mperiod  & Yes & [$-1$, 1] & Gyr  \\
	mskew  & No & [$-1$, 1] & -- \\
	Zfinal  & Yes & [max($-4$, log(Z$_{\rm min}$), $-1.3$]&  -- \\
	\hline
	\multicolumn{4}{l}{$Z_{\odot}$ configuration} \\[3pt]
	\hline
	mSFR  & Yes &[$-3$, 4] & $\rm M_{\odot}/{\rm yr}$ \\
	mpeak  & No & [$-2$, $13.4-t_{\rm lb}$]  & Gyr\\
	mperiod  & Yes & [$-1$, 1] & Gyr  \\
	mskew  & No & [$-1$, 1] & -- \\
	Zfinal  & -- & 0.02 &  -- \\
	\hline
	\multicolumn{4}{l}{b5 SFH configuration} \\[3pt]
	\hline
	m1  & Yes & [$-5$, 4] & $\rm M_{\odot}/{\rm yr}$ \\
	m2  & Yes & [$-5$, 4] & $\rm M_{\odot}/{\rm yr}$ \\
	m3  & Yes & [$-5$, 4] & $\rm M_{\odot}/{\rm yr}$ \\
	m4  & Yes & [$-5$, 4] & $\rm M_{\odot}/{\rm yr}$ \\
	m5  & Yes & [$-5$, 4] & $\rm M_{\odot}/{\rm yr}$ \\
	Zfinal  & Yes & [max($-4$, log(Z$_{\rm min}$), $-1.3$]&  -- \\
	Zstart  & - & max($-4$, log(Z$_{\rm min}$))&  -- \\
	\hline
	\hline
	\multicolumn{4}{l}{Dust parameters in all configurations} \\[3pt]
	\hline
	$\tau_{\rm birth}$  & Yes & [$-2.5$, $-1.5$]& -- \\
	$\tau_{\rm screen}$  & Yes & [$-2.5$, 1] & --\\
	$\alpha_{\rm birth}$ &  No & [0, 4] &  --\\
	$\alpha_{\rm screen}$ & No & [0, 4]  & --\\
	${\rm pow}_{\rm birth}$  & -- & -0.7 & --  \\
	${\rm pow}_{\rm screen}$  & --& -0.7 & --  \\
	\hline
\end{tabular}
\end{table}

Each spectral library analysed in this work is resampled onto the low-resolution BC03 grid, such that fitting is conducted at the same spectral resolution. 

The parameters used for fitting in each configuration are outlined in Table \ref{tab:FreeParameters}. 
The four free dust parameters are applied identically across all implementations, and in addition the fiducial and constant Z implementations have an additional four free SFH parameters and one free metallicity parameter. 
The $Z_{\odot}$ implementation only has an additional four free SFH parameters on top of the free dust parameters, and the b5 SFH implementation has five free SFH parameters and one free metallicity parameter in addition to the free dust parameters. 
When fitting the metallicity for each galaxy, we excluded any templates in the fitting with $Z < 10^{-4}$ to ensure a more even comparison between the different SPLs. 
Note that in our use of \textsc{ProSpect}, we impose no restrictions on the age/metallicity combinations made available by the chosen SPL. 

The optimizer \textsc{Highlander}\footnote{\url{https://github.com/asgr/Highlander}} is used for all SED fitting, as done in recent applications of \textsc{ProSpect} by \citet{thorne2021, thorne2022, thorne2022a, thorne2023}. This tool alternates between the use of the genetic algorithm \textsc{cmaeshpc}\footnote{\url{https://github.com/asgr/cmaeshpc}} and the MCMC (Markov Chain Monte Carlo) CHARM\footnote{Component-wise hit and run metropolis} algorithm within the \textsc{LaplacesDemon}\footnote{\url{https://cran.r-project.org/web/packages/LaplacesDemon/index.html}} package to first find a solution quickly, and then sample over parameter space to characterise local likelihood space and define sampling uncertainty. 
Two CMA$+$CHARM runs with 1000 steps each are used to sample parameter space for each galaxy. Rather than implement a fixed burn-in, all steps with a resulting likelihood $$\rm{LP} < \rm{LP}_{\rm best} - 2\times(\rm{LP}_{\rm median} - \rm{LP}_{\rm best})$$ are discarded, where LP is the log posterior of each step. Typically this discards around 20\% of the chain.

\section{Results}
\label{sec:Results}

SED fitting is such a powerful tool because it is able to provide an expansive range of information about the galaxies being studied. 
To assess the relative performance of the SPLs and the SED fitting configurations, we focus our analysis on the most informative outputs: the quality of fit to the observed SEDs (fitting likelihood), the relative properties of stellar mass and SFR, the derived shape of the mass--metallicity relation (where relevant), the inferred ages, and the ability to accurately return the cosmic star formation history (CSFH). 

Using our fiducial SFH and metallicity history implementation (consistent with the approach outlined in \citealt{bellstedt2020b, bellstedt2021}), we quantify the impact of only changing the underlying SPL on the output galaxy properties. 

\begin{figure*}
	\centering
	\begin{subfigure}[b]{\textwidth}
		\includegraphics[width=180mm]{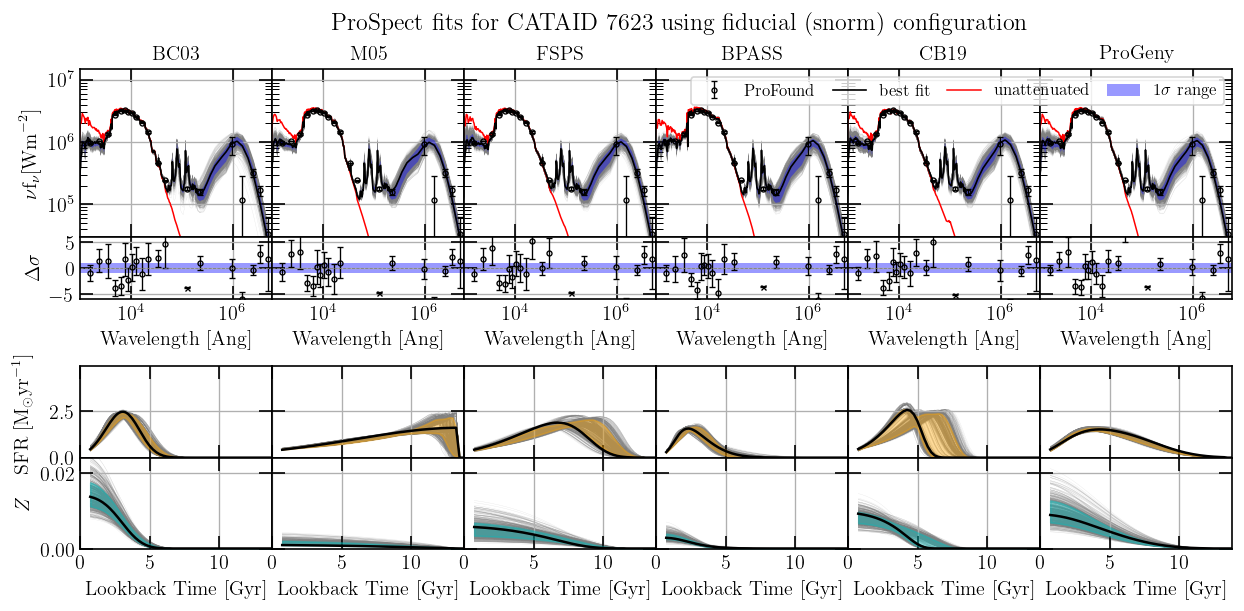}
	\end{subfigure}
	\begin{subfigure}[b]{\textwidth}
		\includegraphics[width=180mm]{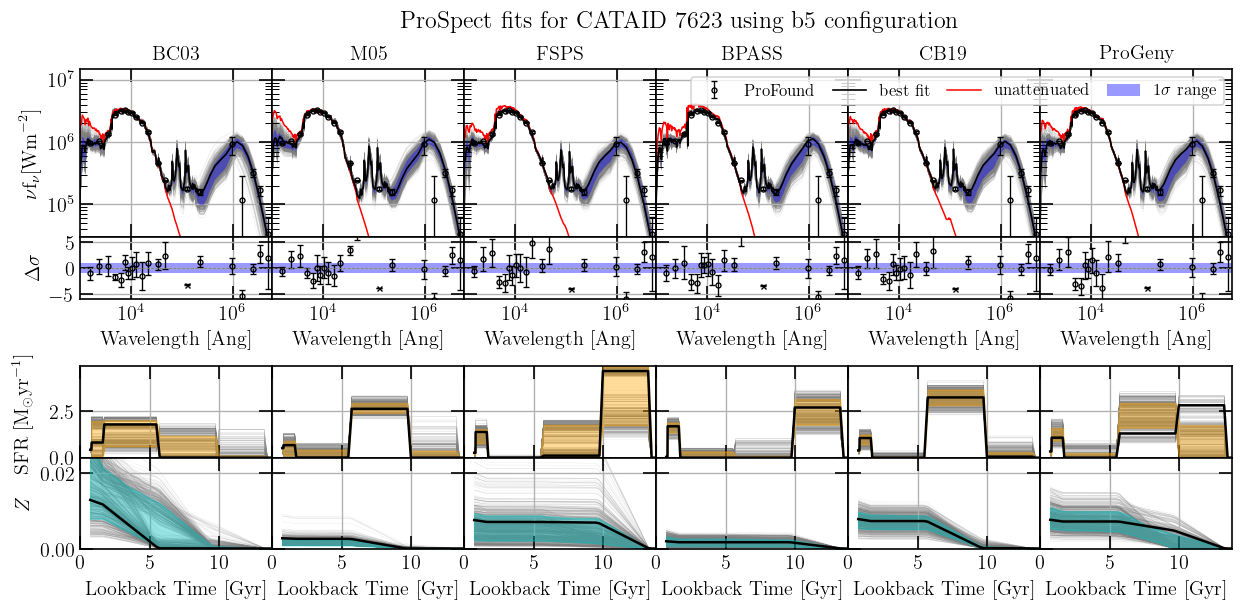}
	\end{subfigure}
	\caption{The \textsc{ProSpect} output for galaxy 7623 with the fiducial configuration (top) and b5 configuration (bottom), for each SPL analysed in this work.   }
	\label{fig:ProSpectExample}
\end{figure*}

Examples of SED fitting outputs for a single galaxy (CATAID 7623\footnote{Detailed information on this source is available through the GAMA Single Object Viewer \url{https://www.gama-survey.org/dr4/tools/sov.php}. }) are provided in Fig. \ref{fig:ProSpectExample}. 
The fiducial configuration has been used in the upper plot of Fig. \ref{fig:ProSpectExample} for each SPL, whereas the b5 configuration has been presented in the lower plot. 
These examples provide an indication of the general quality of fit given the data quality, with the typical fitted range in star formation and metallicity histories. 
Differences in the SFHs are seen when varying the SPLs, but also (unsurprisingly) when changing the SFH parametrisation. It is notable that the metallicity histories are generally less affected by the SFH parametrisation change than the SPL change in this example. 

\begin{figure}
	\centering
	\includegraphics[width=85mm]{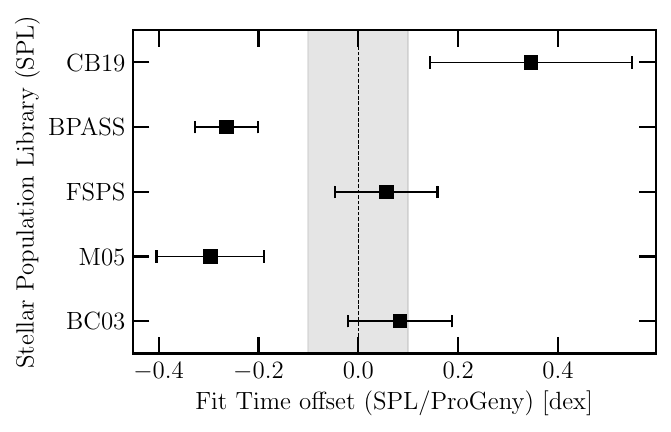}
	\caption{The mean fitting time offsets derived when using different stellar population libraries, as compared with the fitting time estimates derived using \textsc{ProGeny}. }
	\label{fig:FitTimeOffset}
\end{figure}

Due to the often substantial differences between SPLs in terms of temporal and metallicity resolution, the computational requirement of using a different SPL can be substantial. This is reflected in Fig. \ref{fig:FitTimeOffset}, where we compare the amount of time required to fit a galaxy (using the same number of steps in the fitting process) with the varying SPLs (relative to \textsc{ProGeny}). 

We find that the CB19 SPL requires the most computational time (as fitting time scales with the product of the number of age and metallicity bins in the SPL, and CB19 has the greatest number with $221\times15$\footnote{Table 4 from \citet{robotham2024a} presents the $N_{\rm Age}$ and $N_{\rm Z}$ numbers for each SPL. }), with M05, and BPASS requiring 0.6 dex less computation time. M05 and BPASS typically require only 3-5 minutes to fit, \textsc{ProGeny}, BC03 and FSPS require 5-10 minutes of fitting time, whereas CB19 requires up to half an hour for a single fit. 

\begin{figure}
	\centering
	\includegraphics[width=85mm]{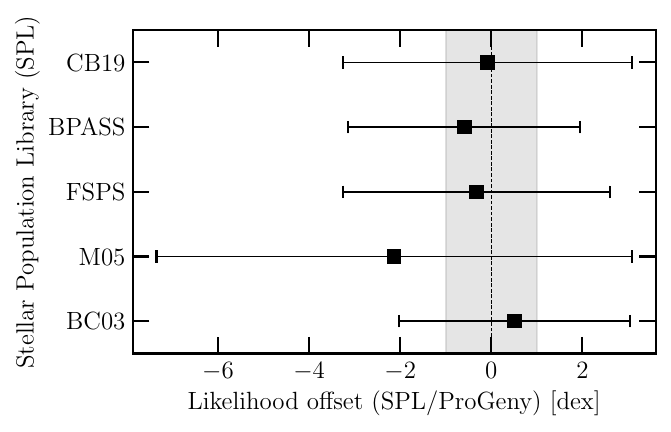}
	\caption{The mean likelihood offsets derived when using different stellar population libraries, as compared with the likelihood estimates derived using BC03hr. On average, BC03hr produces the SED fits with the highest likelihood.  }
	\label{fig:LPOffset}
\end{figure}

The average quality of fit for the SED is shown in Fig. \ref{fig:LPOffset} for each SPL. 
The shaded region is purely for reference, indicating $\pm0.1$ dex (and this shading is used similarly for subsequent plots comparing parameter offsets between SPLs). 
It can be seen that except for M05, for which the resulting SED fits have a slightly lower likelihood, the quality of the resulting fit is comparable. As discussed by \citet{robotham2024a}, M05 differs from the other SPLs in that it most notably lacks the UV upturn\footnote{This is a consequence of the standard M05 version used, which uses red horizontal branch (HB) stars at all metallicities, rather than the hot HB (sometime referred to as extreme HB or eHB) stars at high metallicity, which would produce a UV source for old ages. }. 

\begin{figure}
	\centering
	\includegraphics[width=85mm]{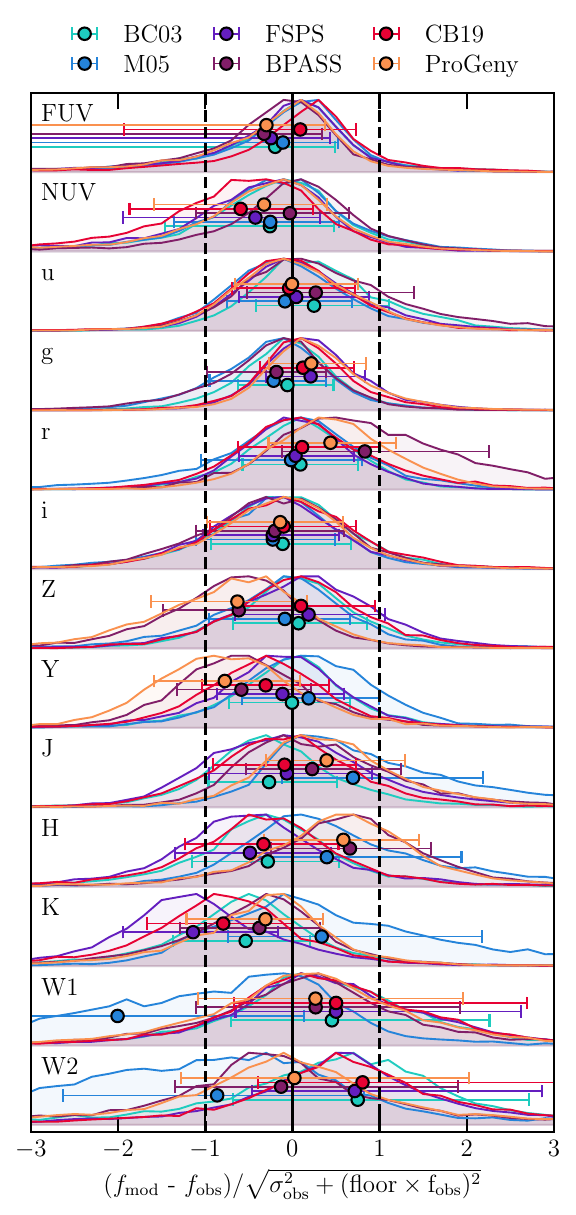}
	\caption{The per-band offset of the modelled photometry from the observed photometry for the whole sample, as modelled using the different SPLs.  }
	\label{fig:PhotometryComp}
\end{figure}

To further assess how the differences in the SPL contribute to modelling offsets across the far-UV to near-IR, we show the distribution of modelled$-$observed photometry divided by uncertainty in each band in Fig. \ref{fig:PhotometryComp}, separately for each used SPL (using the fiducial SED fitting implementation). In each band, the distribution of offsets for the full $\sim$8,000 galaxy sample is shown with a histogram, and the median value with $1\sigma$ range is shown with the error bars. 
We do not show the far-IR in this plot, as it is dominated by dust emission, and therefore not indicative of the SPL behaviour. 
While the quality of fit is in general good (as indicated by the fact that the median values are generally within $\pm 1$), in some bands there is indication of ``over-fitting", evidenced by the fact that the 1$\sigma$ spread of offsets is narrower than $\pm 1$. 
Bands exhibiting this behaviour include $u$, $g$, $r$, and $i$. 
This is potentially an indication that the per-band error floors (determined by \citealt{bellstedt2020b} based on the full GAMA sample, rather than just this local subset) are too high. 

SPL-dependent trends exist in different parts of wavelength space. 
For example, M05 consistently overestimates the flux in the near-infrared ($Y$, $J$, $H$, and $K$ bands), whereas in this same wavelength range, BC03, CB19, and FSPS tend to slightly underestimate the flux. 
In the optical bands, most SPLs tend to produce very consistent SEDs however the exception to this is BPASS and \textsc{ProGeny}, which produce notably more flux in the $r$ band than all other SPLs. 

Using this fiducial SED-fitting implementation, the mean absolute deviation (MAD) of the modelled photometry across FUV to W2 for the BC03, M05, FSPS, BPASS, CB19, and ProGeny SPLs respectively is 0.256, 0.421, 0.337, 0.361, 0.306, and 0.338. 
This means that BC03 produces SEDs most consistent with observed photometry. 
Other SPLs recover the input photometry with similar accuracy, with M05 producing the most discrepant models. 
This is consistent with the global likelihood values presented in Fig. \ref{fig:LPOffset}. 

\subsection{Impact on derived stellar masses}

\begin{figure}
	\centering
	\includegraphics[width=85mm]{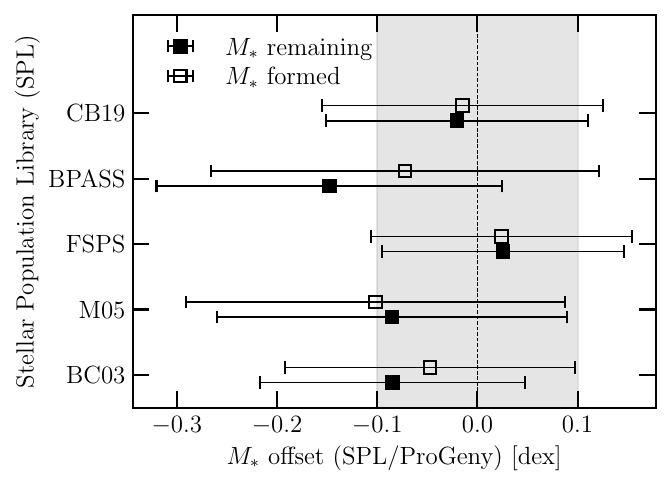}
	\caption{The mean stellar mass offsets derived in the fiducial configuration when using different stellar population libraries, as compared with the stellar mass estimates derived using \textsc{ProGeny}. The remaining mass (what is typically meant by stellar mass) is indicated with solid points, whereas the formed mass is indicated with open points.  }
	\label{fig:StellarMassOffset_SPL}
\end{figure}

The stellar mass of a galaxy is one of the most fundamental properties derived via SED fitting, and therefore this is appropriately the first property we analyse to gauge the impact of SED modelling assumptions. 

The relative deviation in derived stellar masses caused by the choice of SPL is indicated in Fig. \ref{fig:StellarMassOffset_SPL}, where for each SPL we present the mean offset in stellar mass relative to \textsc{ProGeny}. 
FSPS produces the largest stellar masses, followed closely by \textsc{ProGeny} and CB19, whose masses are only $\sim$0.05 dex lower on average. 
The other SPLs tend to produce notably lower stellar masses, offset by $\sim$0.1-0.15 dex on average. 
This behaviour can be directly linked to the underlying stellar recycling fraction implemented in each SPL. 
As demonstrated by \citet{robotham2024a}, the recycled fraction as a function of stellar age varies between the SPLs. 
A larger fraction of the remaining stellar mass is accounted for in SPLs with more stellar remnants. 
FSPS, CB19, and ProGeny all have similar remaining fractions with stellar age, hence explaining why these SPLs tend to recover the most stellar mass. 
\citet{jones2022} compared properties of galaxies derived using CB16 (the early form of CB19) and BPASS, and found the CB16 stellar masses to be greater than those from BPASS by 0.14 dex. This is consistent with the offset that we derive between CB19 and BPASS. 
The open points in Fig. \ref{fig:StellarMassOffset_SPL} compare the total stellar mass formed throughout the history of the galaxy, which does not account for mass loss due to stellar evolution. 
It can be seen that the total mass formed by BPASS is more consistent with the other SPLs than the remaining mass, highlighting that the origin of the mass offset comes from the recycled fraction in BPASS. 
 
\begin{figure}
	\centering
	\includegraphics[width=85mm]{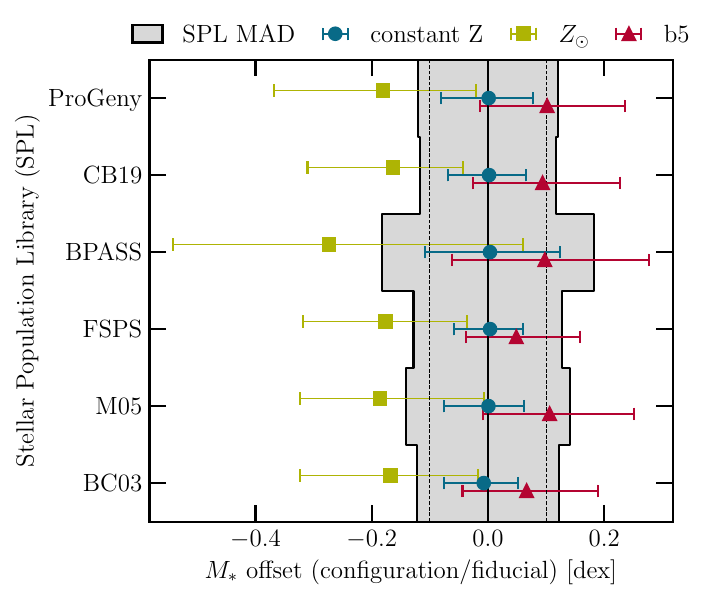}
	\caption{The median stellar mass offsets derived when using different SED fitting configurations, as compared with our fiducial configuration. The shaded region indicates the mean absolute deviation (MAD) of the masses caused by the choice of SPL (the inter-SPL MAD), where a wider grey bar indicates that the SPL produces more discrepant properties.}
	\label{fig:StellarMassOffset}
\end{figure}


Fig. \ref{fig:StellarMassOffset} demonstrates how other assumptions in the SED fitting implementation also impact the stellar mass. 
The coloured points indicate the mean mass offset between each of the $Z_{\odot}$, constant Z, and b5 SFH derived masses as compared with the fiducial stellar masses, for each SPL separately.
The median impact of assuming a constant metallicity (as opposed to the evolving metallicity in the fiducial configuration) is seen to be small, with all SPLs creating a mass offset of at most 0.01 dex. The impact of assuming a fixed solar metallicity however is at times dramatic (depending on the chosen SPL). 
When using BPASS, the impact of this assumption is to reduce the stellar masses by over 0.3 dex, and on average the stellar masses are reduced by around 0.15 dex. The impact of describing the SFH as a stepwise parametrisation is independent of SPL choice however, with a stepwise parametric SFH creating stellar mass values consistently $\sim$0.1 dex higher than a functional parametric SFH. 
This tendency for stepwise parametrisations to produce higher stellar masses than some smooth parametrisations was also demonstrated by \citet{lower2020}. 
Making comparisons of stepwise versus smooth SFHs in general is misleading however, as these are fundamentally linked to the complexity of the suite of SFHs made possible by the parametrisation. 
\citet{carnall2018a} for example showed that stellar masses were overestimated by $\sim$15\% when using a simplistic exponentially declining SFH, whereas by using a double power law the offset was reduced to only 0.02 dex. 
Parametrisations that are more rigid (such as the exponentially declining parametrisation, which is popular because of its simplicity and low number of free parameters) tend to force galaxies to have more similar SFHs, thereby skewing the population properties (hence causing impacts like a 35\% overestimate in stellar masses, as reported by \citealt{ciesla2017}). 

The shaded region in Fig. \ref{fig:StellarMassOffset} shows the mean absolute deviation (MAD) of the SPL values for the fiducial run, represeting the typical mass deviation created through the SPL selection. 
What is notable, is that the mass offsets caused by the constant Z and b5 SED configuration changes is less than the offsets simply caused by the choice of SPL. 

A comparison conducted by \citet{osborne2024} recorded an offset of 0.17 dex between their BC03 and BPASS-derived stellar masses. The difference in their approach is that their metallicity is fixed to solar in each case. As is evident from Fig. \ref{fig:StellarMassOffset_SPL}, BC03 and BPASS produce stellar masses that are only discrepant by $\sim$0.05 dex, whereas Fig. \ref{fig:StellarMassOffset} shows that the impact of fixing metallicity to solar in BPASS is greater than BC03 by 0.13 dex. 
This suggets that the difference presented by \citet{osborne2024} is exacerbated by the choice of metallicity implementation.

\begin{figure*}
	\centering
	\includegraphics[width=180mm]{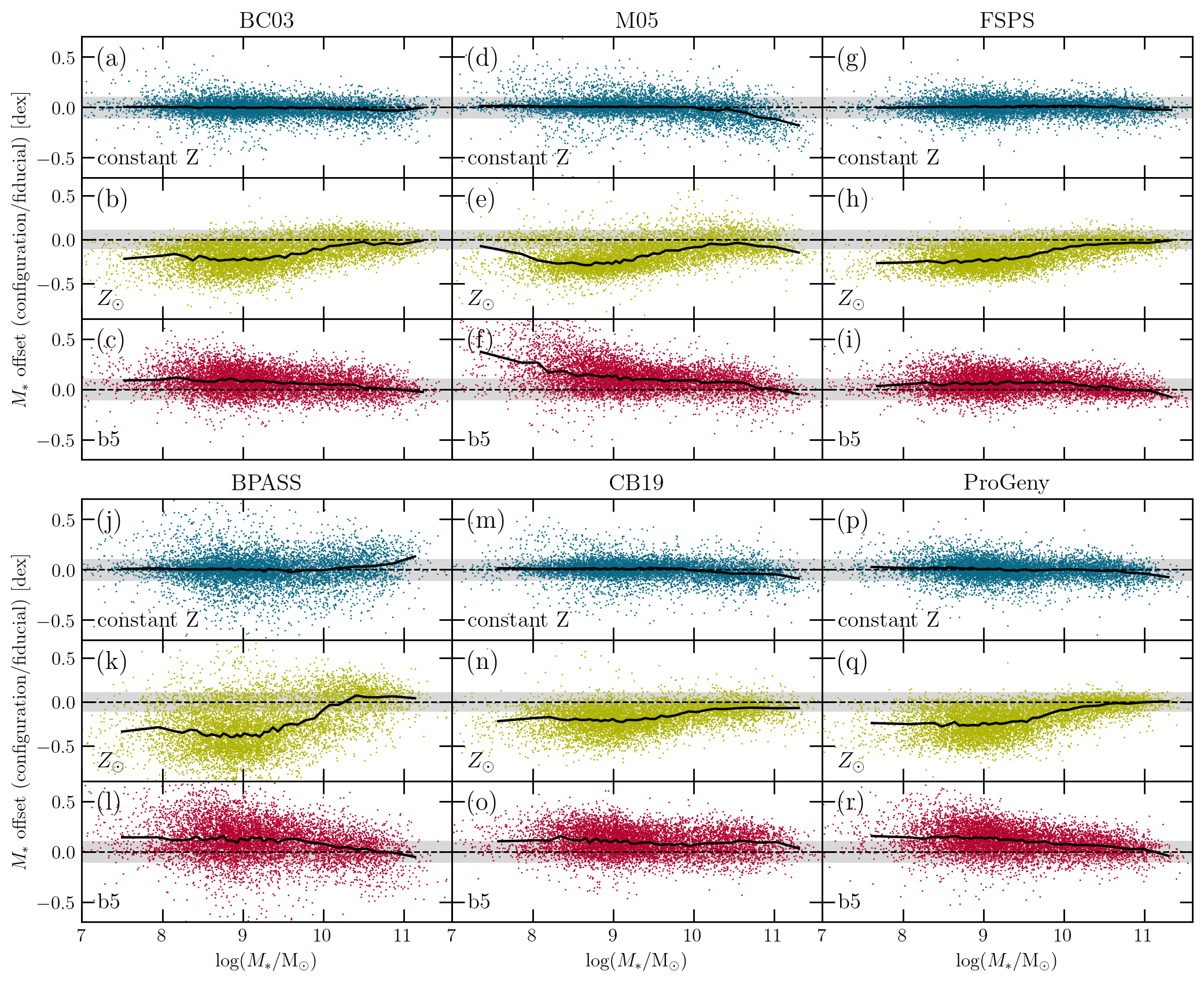}
	\caption{Stellar mass offsets for individual galaxies using different SED configurations, all compared against the fiducial stellar mass estimates. Top panel shows the impact of assuming a constant metallicity, the middle panel shows the impact of fixing the metallciity to solar, and the bottom panel shows the impact of assuming a stepwise parametric SFH. }
	\label{fig:StellarMassOffset_all}
\end{figure*}

While Fig. \ref{fig:StellarMassOffset} shows the median stellar mass impact, it is important to note that these discrepancies are stellar mass dependent (as demonstrated in Fig. \ref{fig:StellarMassOffset_all}). 
While the overall offset created by the constant Z assumption is always less than 0.01 dex, at stellar masses $M_* > 10^{10.5} \rm{M}_{\odot}$ the values are overestimated by 0.06 dex in BPASS, and underestimated by just over 0.04 dex in both CB19 and \textsc{ProGeny}. 
Subpanels (p)--(r) of Fig. \ref{fig:StellarMassOffset_all} show the offsets of individual galaxies using the \textsc{ProGeny} SPL for each SED configuration, as compared with the fiducial implementation. 
The impact of ignoring the presence of a relation between a galaxy's stellar mass and metallicity is evident in sub-panel (q), where the impact of fixing the metallicity to solar is a very strong mass dependence of the stellar mass offset. 
At low stellar masses, galaxies are forcibly modelled with metallicity values that are too large, forcing the model to produce younger stars to create a sufficiently blue SED. 
This results in a lower mass-to-light, and therefore the masses are underestimated. 
At high stellar masses however the assumption of a solar metallicity is more realistic, and hence the stellar mass offset is minimal. 
Sub-panel (r) of Fig. \ref{fig:StellarMassOffset_all} also shows that while the overall effect of a stepwise versus functional parametrisation is to overestimate the stellar masses by around 0.1 dex in \textsc{ProGeny}, at the highest masses the values are actually consistent. 
These trends are qualitatively similar for all SPLs except BPASS, where the impact of a constant metallicity is reversed at higher stellar masses (see sub-panel (l) of Fig.  \ref{fig:StellarMassOffset_all}). 

We highlight that the trends recovered in subpanels (b) and (c) of Fig. \ref{fig:StellarMassOffset_all} for BC03 are exactly consistent with those presented in fig A3 by \citet{bravo2022} where similar testing was conducted to assess the impact of evolving metallicity and stepwise parametric SFHs.

\subsection{Impact on derived SFR}

\begin{figure}
	\centering
	\includegraphics[width=85mm]{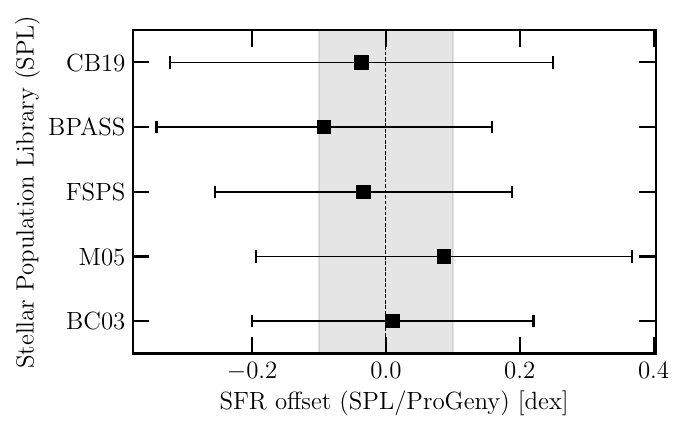}
	\caption{The mean SFR offsets derived when using different stellar population libraries, as compared with the SFR estimates derived using BC03hr. }
	\label{fig:SFROffset_SPL}
\end{figure}

Presenting the results from just the fiducial configuration, we compare the impact of selecting different SPLs on the derived star formation rates (SFRs) relative to the \textsc{ProGeny}-derived SFR in Fig. \ref{fig:SFROffset_SPL}. 
Within \textsc{ProSpect}, the SFR is taken to be the averaged value of the star formation history in the most recent 100 Myr. 
Values are generally consistent within $\pm0.1$ dex, where M05 produces the highest values, and BPASS produces the lowest.  

\begin{figure}
	\centering
	\includegraphics[width=85mm]{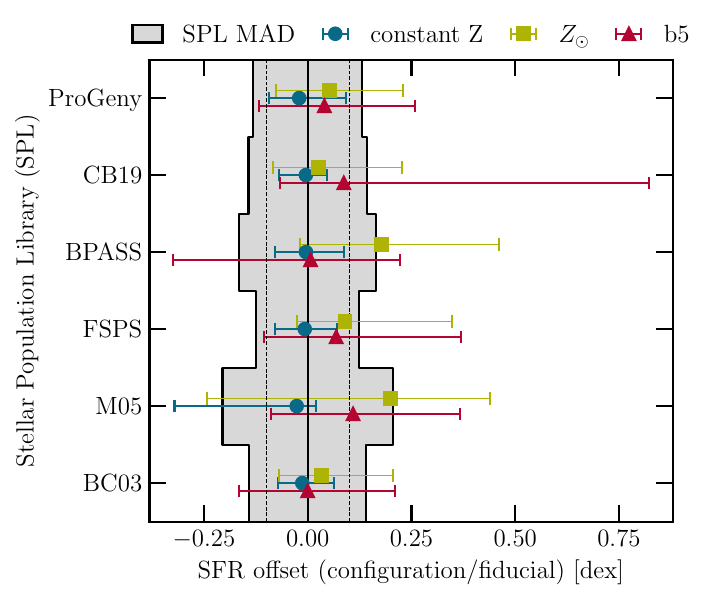}
	\caption{The median SFR offsets derived when using different SED fitting configurations, as compared with our fiducial configuration. The shaded region indicates the inter-SPL MAD. All SFR measurements less than $10^{-6}\,\rm{M}_{\odot}\rm{yr}^{-1}$ are shifted to $10^{-6}\,\rm{M}_{\odot}\rm{yr}^{-1}$ for the sake of this comparison (as the SFR measurements derived using \texttt{massfunc\_snorm\_trunc} can be arbitrarily small). }
	\label{fig:SFROffset}
\end{figure}

The relative impact of different \textsc{ProSpect} implementation configurations as compared with the impact of SPL selection is then shown in Fig. \ref{fig:SFROffset}. 
As in Fig. \ref{fig:StellarMassOffset}, the coloured points represent the offset caused by the SED fitting configuration relative to the fiducial configuration, whereas the dark shaded region represents the mean absolute deviation (MAD) of the SPL values for the fiducial run. 
For runs where the coloured point lies outside of the shaded region, this is an indication that the SED configuration has a greater impact on the SFR measurements than the choice of SPL. 
The SPL choice is the dominant impact on SFRs in almost all cases. 
The only exceptions to this are the $Z_{\odot}$ runs conducted with BPASS and M05, where the resulting SFRs are measured to be around 0.2 dex higher than the fiducial implementation, which is on par with the typical impact from SPL choice. 

\begin{figure}
	\centering
	\includegraphics[width=85mm]{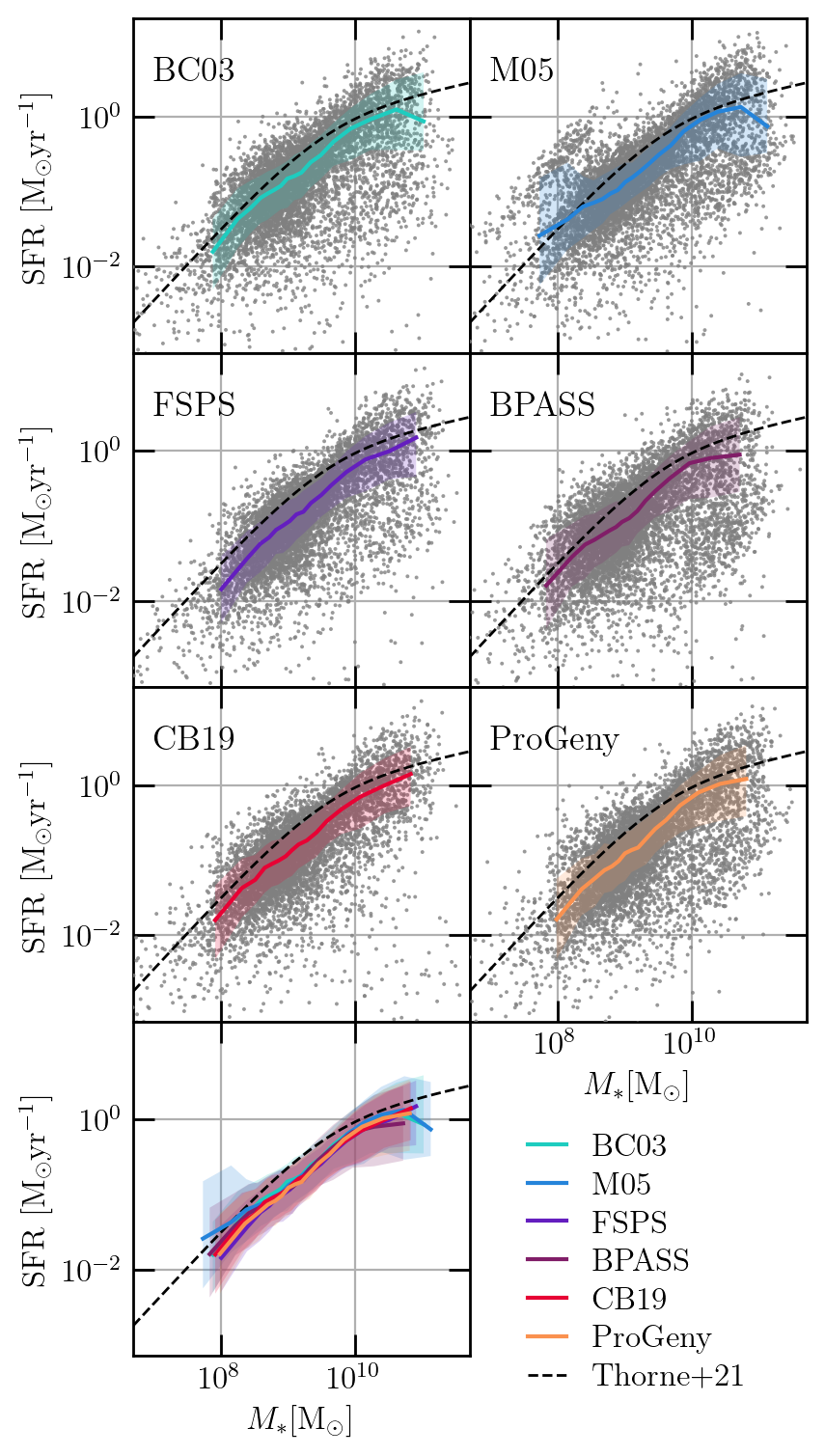}
	\caption{The derived stellar mass -- SFR relation when using different stellar population libraries. }
	\label{fig:SFMS}
\end{figure}

A parameter space that has been extremely well studied in the literature is the stellar mass -- SFR plane, which presents the co-called Star Formation Main Sequence (SFMS). 
Fig. \ref{fig:SFMS} shows how the fiducial implementation produces varying SFMS derivations using the six SPLs. 
The recovery of the SFMS is generally consistent, with a very similar normalisation, and a similar bending of the MS at high masses. This is reflected by the coloured lines, which represent the running median of all galaxies that have SFRs greater than 1 dex below the SFMS fit by \citet{thorne2021}. While these lines are not meant to be a recovery of the SFMS, they provide a straightforward way of comparing the overall behaviour in the bottom panel. There is a population of high-mass ``quenching" galaxies that sit just below the MS in all outputs except that of CB19. 
While none of the galaxies occupying this region ever appear on the main sequence itself (suggesting that the constraint of this lower SFR is real), there is a tendency for them to scatter to the more fully quenched position below the plot.
There is also a population of low-mass ``bursting" galaxies produced by M05, however the fact that they are not recovered by any other SPL suggests that the separation to higher SFRs is an artefact of the fitting when using M05.
The other SPLs do recover these sources consistently as highly star-forming galaxies, in that they either sit directly on or above the main sequence defined by \citet{thorne2021}. 
Finally, we note that the derived SFMS all consistently show greater scatter in the b5 SFH implementation than the fiducial implementation, highlighting the consequence of the freedom in the present-day star formation rate when using stepwise bins in the SFH implementation.

\subsection{Impact on the derived metallicity}

Metallicity is often treated as a nuisance parameter in SED fitting, required to retrieve galaxy properties like stellar mass and SFR. 
However with more complex SFHs and the implementation of physically motivated metallicity evolution, it is increasingly possible to produce realistic estimates of galaxy metallicity using SED fitting \citep[as demonstrated by][]{bellstedt2021, thorne2022a}. 
We therefore show the impact of SPLs and SED fitting configurations on these estimates in this section. 

\begin{figure}
	\centering
	\includegraphics[width=85mm]{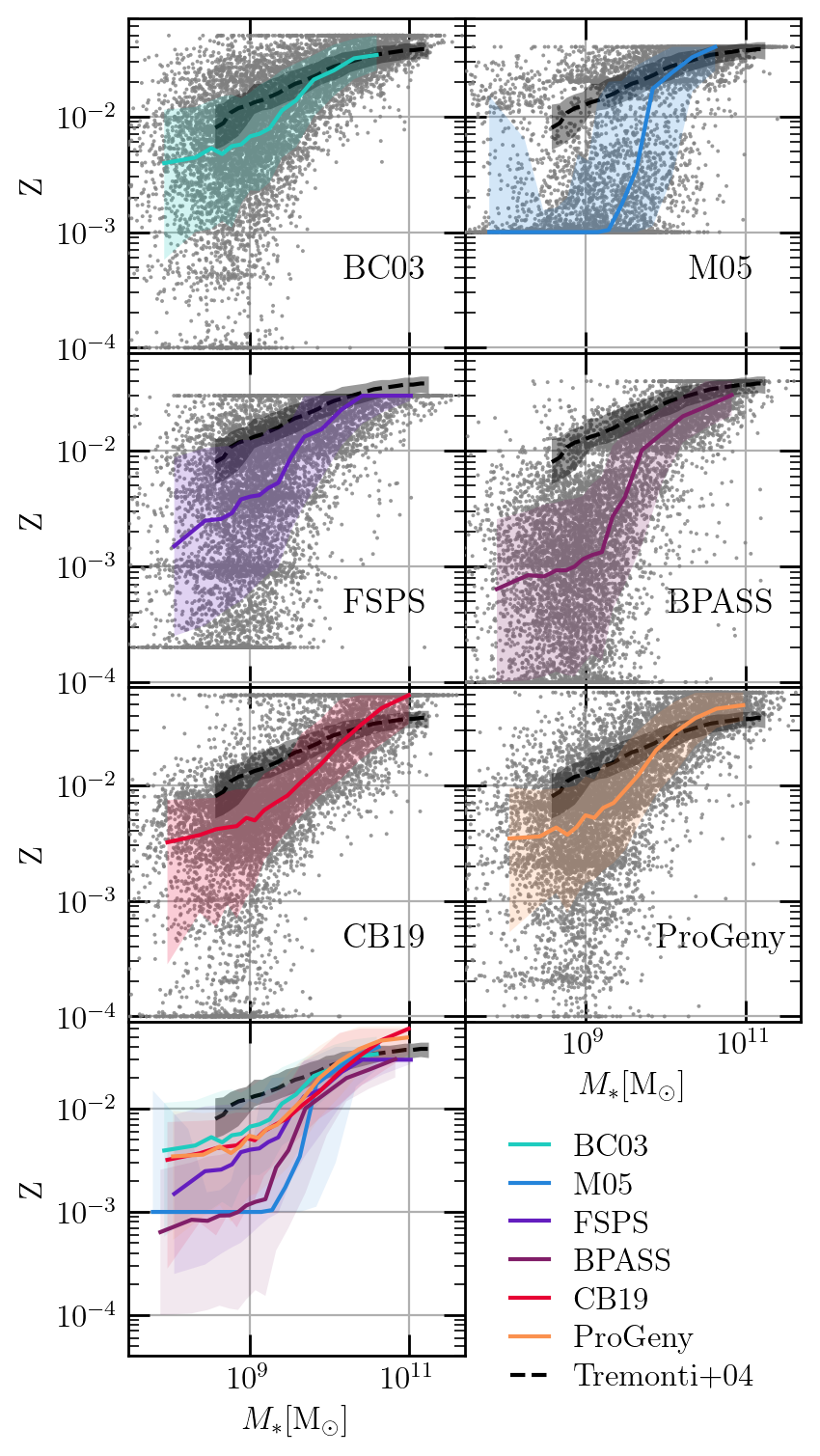}
	\caption{The Mass--Metallicity relation as derived when using different stellar population libraries. The observational mass--metallicity relation from SDSS as derived by \citet{tremonti2004} is overplotted on each panel for reference. The bottom panel compared the mass--metallicity relations inferred from each SPL.  }
	\label{fig:MZR_fiducial}
\end{figure}

The derived mass--metallicity relations using different SSPs are given in Fig. \ref{fig:MZR_fiducial}. Each panel represents the output derived using a different SPL, where values for individual galaxies are indicated as grey points, with the running median shown in the coloured line. 
The inherent limit in metallicity range provided by each SPL can be seen in this figure, where \textsc{ProGeny} has the widest range in possible metallicity values, whereas M05 has the most truncated range. 
For a comparison with observations of direct metallicity measurements, we show the Mass -- Metallicity Relation (MZR) from \citet{tremonti2004} as the dashed black line in each panel. 
The bottom right subpanel compares the median MZR derived by all SPLs. 
The overall shape for each SPL is similar, recovering the classic increase of metallicity with stellar mass, where there is a break in the relation at $\sim 10^{10}\rm{M}_{\odot}$, and a flattening of the curve at the highest stellar masses. 

It is notable that the low-mass values for metallicity are recovered quite differently for the various SPLs. BC03, FSPS, CB19 and \textsc{ProGeny} all have lower median values of metallicity of $\sim3\times 10^{-3}$. 
Metallicity values for low-mass galaxies in M05 and BPASS are substantially lower, with $<10^{-3}$. 
For M05 this is largely driven by the fact that low-mass galaxies are hitting the lower bound of the available templates, while BPASS seems to genuinely prefer fitting lower metallicity values. 
The SPL that produces a MZR most similar in shape to that of \citet{tremonti2004} is BC03. 
It is worth noting that \citet{tremonti2004} use the BC03 library in the derivation of their metallicity values, so this agreement is perhaps not surprising (while acknowledging that spectrally derived and photometrically derived metallicities are not expected to be the same). The general agreement between photometrically derived and spectrally derived trends using the same SPL and very different techniques is reassuring, however.

\begin{figure}
	\centering
	\includegraphics[width=85mm]{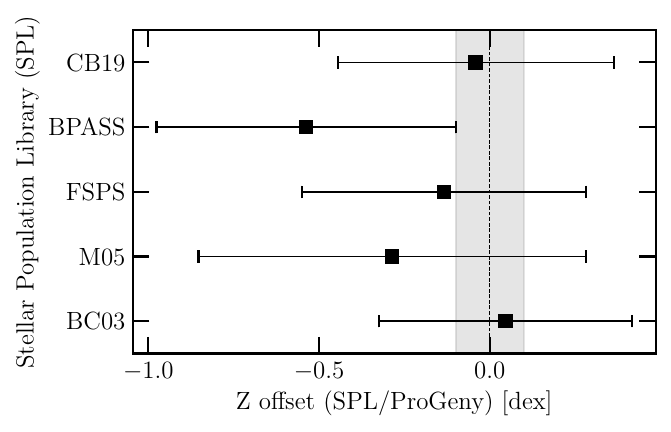}
	\caption{The mean Zfinal offsets derived when using different stellar population libraries, as compared with the Zfinal estimates derived using \textsc{ProGeny}. }
	\label{fig:ZfinalOffset_SPL}
\end{figure}

The overall metallicity values are compared at the population level in Fig. \ref{fig:ZfinalOffset_SPL}. 
Given the complexity in deriving metallicity values from photometry alone (as demonstrated by the overall offsets in the derived MZRs in Fig. \ref{fig:MZR_fiducial}), it is unsurprising that the variation in derived metallicity parameters in Fig. \ref{fig:ZfinalOffset_SPL} is much greater than the variation in stellar mass or SFR values, with the median offset between \textsc{ProGeny} and BPASS around 0.5 dex.

\begin{figure}
	\centering
	\includegraphics[width=85mm]{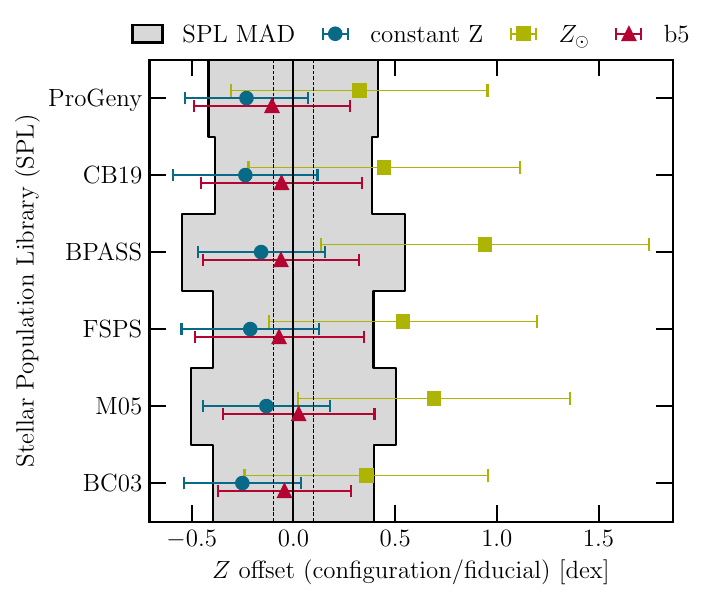}
	\caption{The mean Zfinal offsets derived when using different SED fitting configurations, as compared with our fiducial configuration. The shaded region indicates the inter-SPL MAD. }
	\label{fig:ZfinalOffset}
\end{figure}

Looking beyond only the impact of SPL on derived metallicities, Fig. \ref{fig:ZfinalOffset} shows the impact of SED fitting implementations. 
Given that two of the tested configurations explicitly change the metallicity implementation, it is unsurprising that these have a substantial impact.
As expected, the most severe of these comes from assuming a fixed solar metallicity, as most galaxies in the sample would be expected to have sub-solar metallicities. 
The impact of assuming a constant metallicity means that earlier stars are modelled with a higher metallicity than expected, requiring that the latest stars formed in a galaxy are modelled with a lower metallicity than expected. 
This is evident in Fig. \ref{fig:ZfinalOffset}, where it can be seen that the constant Z metallicities are underestimated by around 0.2 dex. 
Note that the average metallicity across the metallicity history (ZH) of a galaxy in our fiducial evolving metallicity model is typically 0.3 dex lower than the fitted \texttt{Zfinal} value, (around \texttt{Zfinal}/2), so the constant Z metallicity is more a reflection of this than the present-day metallicity. 

Interestingly, the assumed SFH parametrisation also has an impact, although it seems that this is linked mostly to the relative offsets derived from the SFRs themselves (the trend in metallicity of the red points in Fig. \ref{fig:ZfinalOffset} is very similar to that of the red points in Fig. \ref{fig:SFROffset}). 
Note however that the effects of constant Z and the b5 SFH parametrisation are all sub-dominant to the impact of the selected SPL itself (marked in Fig. \ref{fig:ZfinalOffset} with the grey shaded region).

\subsection{Impact on the derived SFHs}

\begin{figure}
	\centering
	\includegraphics[width=85mm]{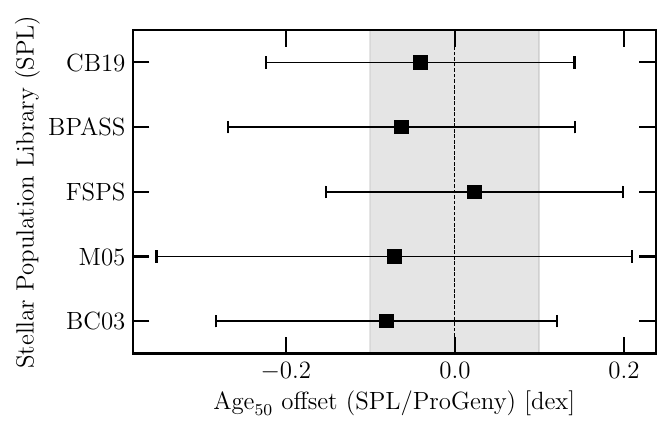}
	\caption{The mean Age$_{50}$ offsets derived when using different stellar population libraries, as compared with the Age$_{50}$ estimates derived using \textsc{ProGeny}. }
	\label{fig:HalfMassAgeOffset_SPL}
\end{figure}

\begin{figure}
	\centering
	\includegraphics[width=85mm]{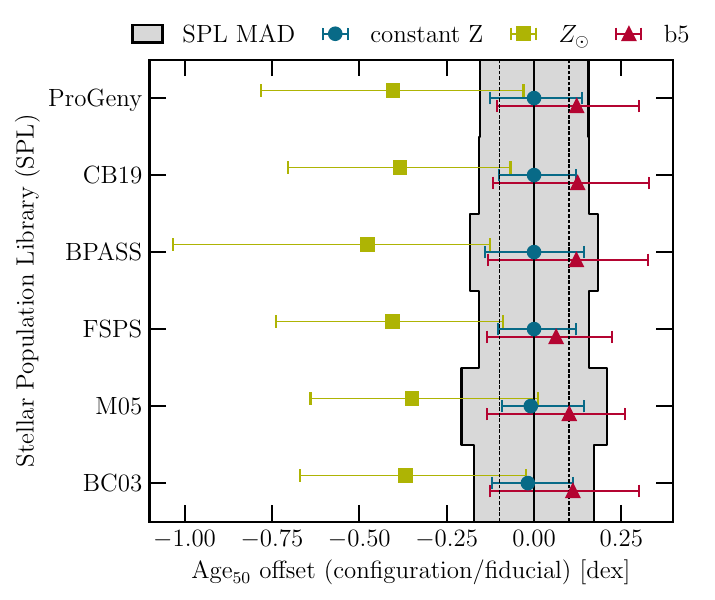}
	\caption{The mean Age$_{50}$ offsets derived when using different SED fitting configurations, as compared with our fiducial configuration. The shaded region indicates the inter-SPL MAD. }
	\label{fig:HalfMassAgeOffset}
\end{figure}

A simple way of comparing fitted SFHs of galaxies is to assess the inferred ages, computed as the age at which half the galaxy's stellar mass is formed (Age$_{50}$). 
This can also be thought of as the median star formation age. 
Fig. \ref{fig:HalfMassAgeOffset_SPL} presents the median offsets of these derived ages when using different SPLs. On averge, Age$_{50}$ estimates all agree within around 0.1 dex. 
The impact on Age$_{50}$ by SED fitting implementation is then shown in Fig. \ref{fig:HalfMassAgeOffset}. 
As with other global parameters presented in this work, the impact of the $Z_{\odot}$ implementation is substantial, with ages underestimated by around 0.4 dex. 
The b5 SFH implemenation causes an overestimate of the ages by around 0.1 dex, and on average the constant Z implementation produces no bias in the Age$_{50}$. 

\begin{figure*}
	\centering
	\includegraphics[width=180mm]{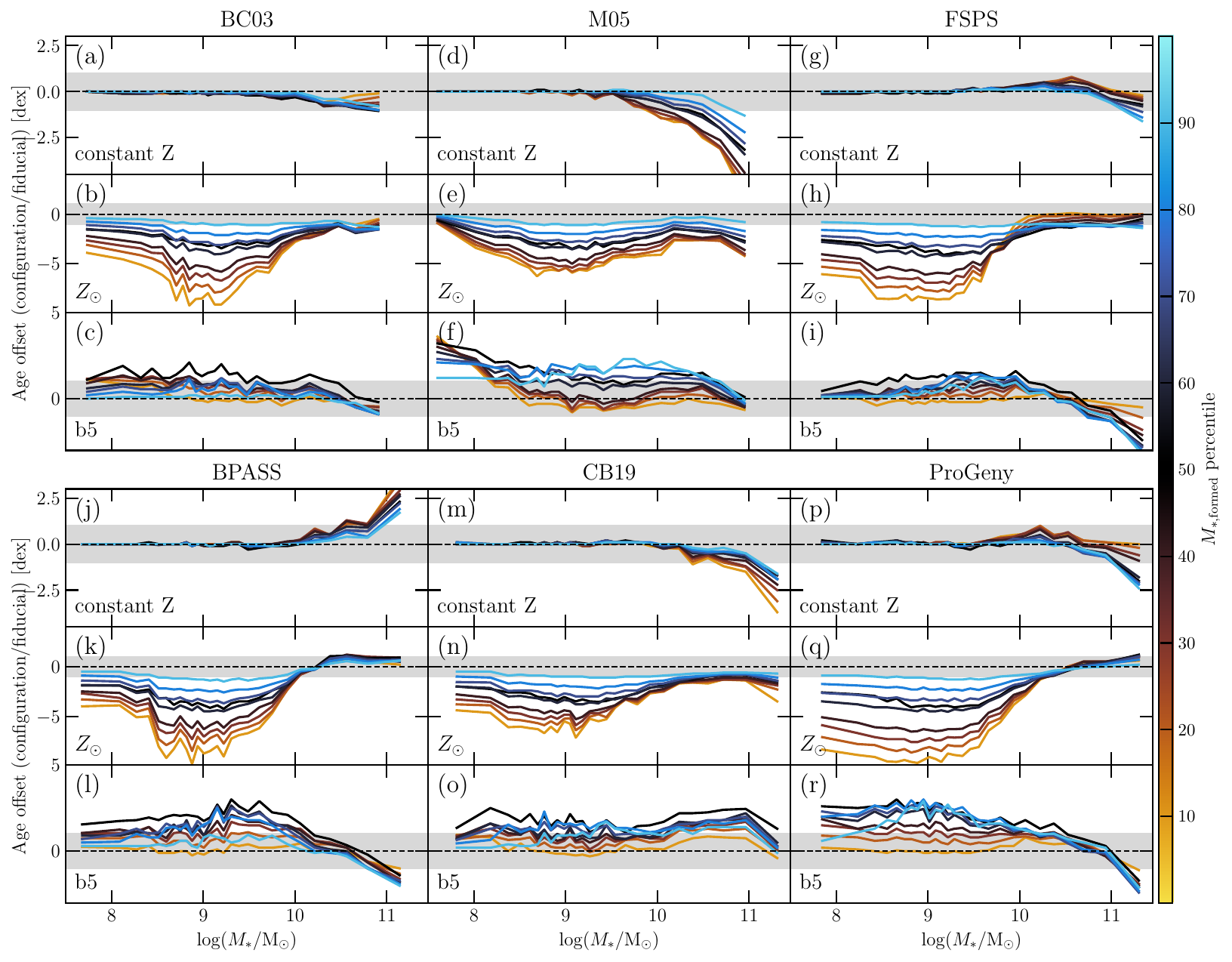}
	\caption{Median age offsets from the fiducial implementation as a function of stellar mass, at varying $M_{*, \rm formed}$ percentiles, using the \textsc{ProGeny} SPL.  }
	\label{fig:AgeOffsetEvolution_all}
\end{figure*}

While the Age$_{50}$ provides a simplified way of condensing SFH information, it does not convey the full picture. 
Fig. \ref{fig:AgeOffsetEvolution_all} shows the median offset in ages computed at different $M_{*, \rm formed}$ percentiles as a function of galaxy stellar mass when using the different SED fitting implementations. 
As with Fig, \ref{fig:StellarMassOffset_all}, it is divided into six panels, showing the impact of the individual SPLs.  

For a detailed assessment of the behaviour, we first focus on sub-panels (p)--(r), showing the impacts using the \textsc{ProGeny} SPL. 
In each sub-panel, the black line shows the half-mass age (Age$_{50}$). Redder lines show the ages computed for lower mass quantiles (corresponding to younger ages, at earlier times) and bluer lines show the ages computed for higher mass quantiles (corresponding to older ages, at later times). 
Sub-panel (p) shows the relative offset in ages between the constant Z implementation, and the fiducial implementation. For galaxies with $M_* < 10^{10} \rm{M}_{\odot}$ there is no offset in ages, which explains the global trend in Fig. \ref{fig:HalfMassAgeOffset}. However, for the most massive galaxies, a deviation is seen. In \textsc{ProGeny}, even though the ages at which galaxies first start forming stars are consistent, the final stars are formed much later in the constant Z implementation than the fiducial implementation, indicating a longer period of star formation. 
Even though the population-averaged age statistics are consistent, the fact that a bias exists for the most massive galaxies has consequences in terms of any mass-weighted analysis (such as the inferred cosmic star formation history, see the next Section). 

Sub-panel (q) focuses on the relative offset between the $Z_{\odot}$ implementation and the fiducial implementation. The trends seen here are reversed to the top panel, where the average behaviour is a very large offset towards younger ages. At the massive end, however, this offset reduces substantially, indicating that the SFHs between the $Z_{\odot}$ implementation and the fiducial implementation are fairly consistent at $M_* > 10^{10.5} \rm{M}_{\odot}$. 
Both these top two panels highlight that the metallicity dependence of the studied population (as evidenced by the mass--metallicity relation) has the capacity of introducing a notable bias in the SFHs when a physical metallicity evolution is not taken into account. 

Finally, sub-panel (r) shows the relative offset between the b5 SFH implementation and the fiducial implementation. Again, the trend is very mass-dependent, with the lower-mass galaxies having older ages in b5 SFH than in the fiducial configuration. In particular, the offset is greater for the later stars formed, indicating that the SFHs are more truncated in time in b5 SFH than the fiducial implementation. 
The mass dependence of the offset continues at the highest masses, where SFHs are recovered to be younger in b5 SFH at $M_* > 10^{11} \rm{M}_{\odot}$. 

Qualitatively, the behaviour presented for \textsc{ProGeny} is also present in FSPS (sub-panels (g)--(i) of Fig. \ref{fig:AgeOffsetEvolution_all}), however this is not true for all SPLs. When implementing a constant metallicity in BPASS, for example (sub-panel (j) of Fig. \ref{fig:AgeOffsetEvolution_all}), the duration of the SFH also expands at higher masses, however instead of being biased to more recent, it is biased toward older SFHs. This can be identified through the coloured lines moving upwards in the plot, as opposed to downwards as seen for \textsc{ProGeny}.

Different behaviour again is noted for the CB19 SPL (sub-panels (m)--(o) of Fig. \ref{fig:AgeOffsetEvolution_all}). Like \textsc{ProGeny} the median offsets in the constant Z implementation are biased downwards at high mass, however the colour trend is reversed (a feature that also occurs in the M05 SPL). Not only are the SFHs skewed towards later times, but the SFH is truncated in terms of duration (such that the early star formation occurs later). In general, the consequences of the $Z_{\odot}$ and b5 SFH implementation are fairly consistent. 

\subsubsection{Manifestation of the age--metallicity degeneracy}
\label{sec:AgeMetallicity}

\begin{figure*}
	\centering
	\includegraphics[width=180mm]{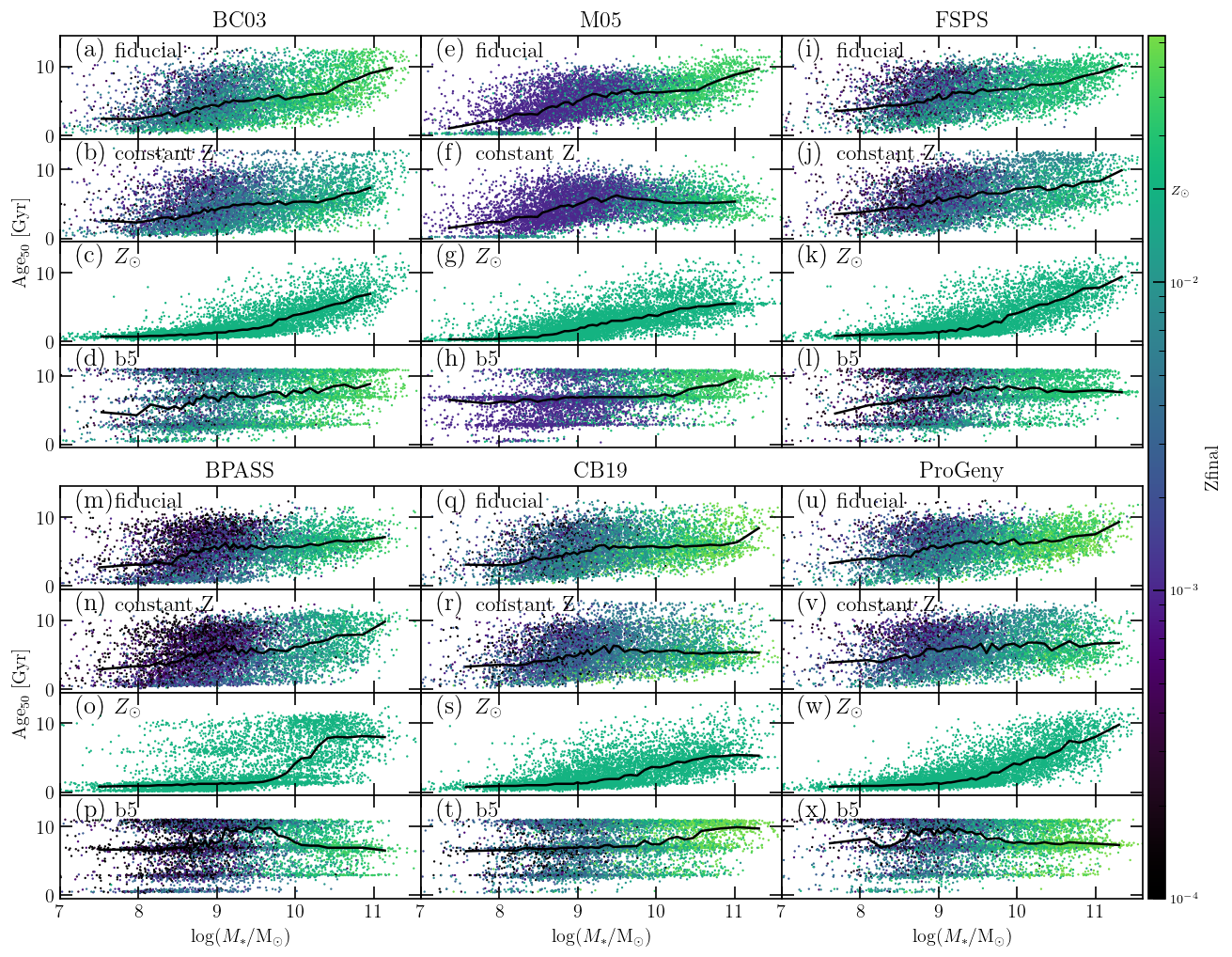}
	\caption{The distribution of half-mass ages with stellar mass in each SED fitting configuration, with individual SPLs selected. Individual points are coloured by the Zfinal value, and the solid black line in each subpanel shows the running median.} 
	\label{fig:Age_Mass_Z}
\end{figure*}

While we have studied the impact of set fitting choices on the metallicity and the ages separately, the known age--metallicity degeneragy \citet[ADD CITATIONS HERE][]{} means that it is difficult to study these impacts in isolation. Fig. \ref{fig:Age_Mass_Z} presents the half-mass age as a function of the galaxy stellar mass for each configuration tested, coloured by the final metallicity value \texttt{Zfinal} to better compare these outputs. The overall conclusion from our analysis is that the degree to which the age-metallicity degeneracy falsely drives the fitted parameters changes depends on the SED fitting configuration, which we will elaborate on in this section. 

The most immediate impact of the age--metallicity degeneracy noted, is that when the metallicity is set to be fixed in the SED fitting, then the only remaining lever that can be modified to fit the observed colours is age. This can be seen in the $Z_{\odot}$ subpanels, where the resulting age is directly correlated with the galaxy mass (a consequence of the fact that galaxies at higher masses tend to be redder). Interestingly while this behaviour is consistent across the different SPLs, BPASS is a notable exception, as seen by the different appearance of panel (o).
The likely cause of this behaviour is the notably different UV flux presence in the BPASS templates. Fig 11 of \citet{robotham2024a} compiles the templates at solar metallicity across a range of ages for each of the SPLs studied in this work. At every age, the BPASS template is an outlier within the UV. At lower ages the UV flux is higher, and at older ages the UV flux is lower. This may well explain the slightly bimodal behaviour in panel (o) of Fig. \ref{fig:Age_Mass_Z}, where at fixed $Z_{\odot}$ \textsc{ProSpect} tends to choose either younger or older ages, depending on the UV flux in a galaxy.  

When metallicity is allowed to vary, then at a given stellar mass the range in derived ages becomes greater. This represents a relaxing of the consequence of the age--metallicity degeneracy. In the constant Z implementation, the sample-wide range in recovered ages tends to be lower than in the $Z_{\odot}$ implementation, showing that the freedom in metallicity is a more dominant lever when trying to reproduce the observed colours of a galaxy. The fiducial implementation, with its evolving metallicity prescription, presents a compromise between these two trends. The variation of age across each mass bin is still broad, highlighting that the age lever is not dominating, however the sample-wide range in recovered ages is also broader, highlighting that the fitted metallicity is facilitating the more extreme age values that would be hampered by a constant metallicity implementation. 
At the highest masses, the recovered ages are older in the fiducial panel than the constant Z panel, and have a higher \texttt{Zfinal} value. This is because the oldest stars can form at a low metallicity in the fiducial case, allowing \textsc{ProSpect} to favour the oldest ages. Again, a notable exception to this trend is BPASS, where the highest mass galaxies in the constant Z implementation in panel (n) actually have older ages than the same galaxies in the fiducial implementation in panel (m). We believe this can again be attributed to the UV flux behaviour in BPASS, where for the highest metallicities, BPASS consistently produces more UV flux than other SPLs (see fig B18 in \citealt{robotham2024a}). This leads \textsc{ProSpect} to disfavour high-metallicity outputs in the fiducial implementation, thereby limiting the ages somewhat. 

This analysis highlights that when limited flexibility is afforded in an SED fitting implementation, the age--metallicity degeneracy will drive the resulting properties. When afforded greater flexibility and more physical implementations however, the impact of this degeneracy is minimised. 

Finally, evident from the b5 subpanels in Fig. \ref{fig:Age_Mass_Z} is the discretisation in age, reflecting favoured half-mass age values when using the stepwise parametric function for the SFH. This will occur when a large fraction of the stellar mass in a galaxy is formed within a single time bin, which because of the fixed width of each bin results in discretised half-mass ages. The existence of these artefacts in the age derivation mean that the impact of the age-metallicity degeneracy here is more difficult to discern.

\subsection{Impact on Derived Cosmic Star Formation History}

\begin{figure*}
	\centering
	\includegraphics[width=180mm]{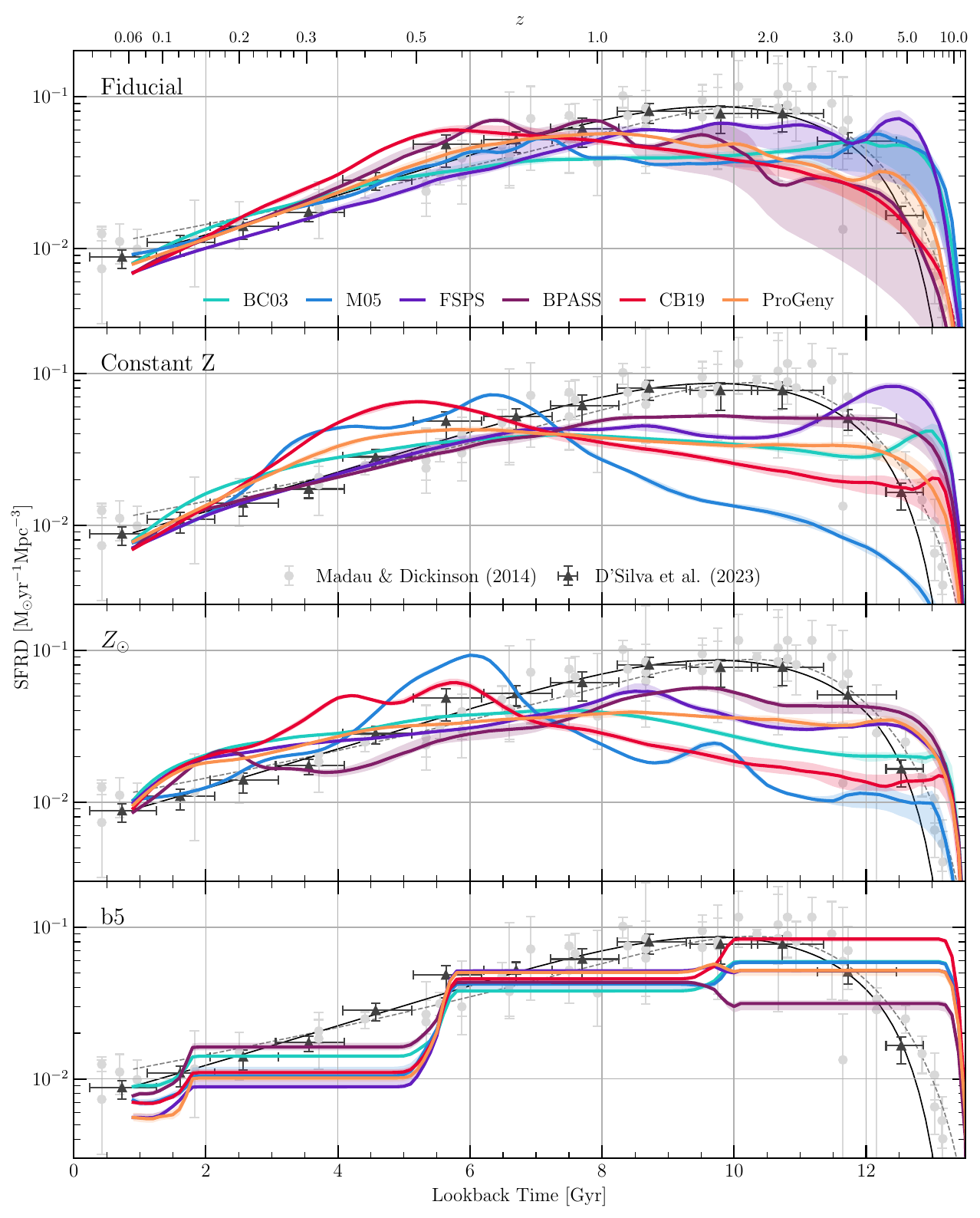}
	\caption{The CSFH as derived when using different stellar population libraries with the different SED fitting configurations shown relative to the compilation of observations from \citet{madau2014} and the more recent derivation from \citet{dsilva2023}.}
	\label{fig:CSFH_all}
\end{figure*}

We can use the volume-limited nature of our GAMA sample to compute the CSFH using the individual SFHs in the same manner as outlined by \citet{ carnall2019, leja2019a, bellstedt2020b, bellstedt2024}, as a manner of assessing the systematic bias of the SFHs. 
The CSFH is well understood from high-$z$ observations from a number of works \citep[including][]{madau2014, driver2018, dsilva2023}, and therefore any deviation in the forensic\footnote{`Forensic' here refers to the process of constructing a historical property (the CSFH) from present-day galaxies.} reconstructed CSFH and the observed one points to underlying age biases in the SED fitting process. 

The methodology of computing the CSFH from individual SFHs is identical to that outlined by \citet{bellstedt2024}. Briefly, each individual SFH is shifted to be in lookback time units (necessary because individual SFHs are computed relative to the epoch of the galaxy), and then all SFHs are summed to produce a total SFH. Incompleteness in the sample is accounted for by scaling the SFH contributions of individual galaxies with $V_{\rm max} < V_{\rm sample}$ by a volume correction of $V_{\rm sample}/V_{\rm max}$, using the \textsc{ProSpect}-derived $V_{\rm max}$ values. If $V_{\rm max}\geq V_{\rm sample}$, then no correction is applied. Finally, the summed SFH is divided by the total volume of the sample to derive the star formation rate density (SFRD) with lookback time. We only present the CSFH at $z>0.06$, as below $z=0.06$ a changing number of galaxies contributes to the CSFH, resulting in numerical artefacts. 

The top panel of Fig. \ref{fig:CSFH_all} presents the derived CSFH using the fiducial implementation for all six SPLs analysed in this work, as compared with the compilation of measurements that derive the star formation rate density using either UV or FIR luminosity functions by \citet{madau2014}\footnote{Measurements included in this compilation come from a suite of studies: \citet{sanders2003, takeuchi2003, wyder2005, schiminovich2005, dahlen2007, reddy2009, robotham2011, magnelli2011, magnelli2013, cucciati2012, bouwens2012, bouwens2012a, schenker2013, gruppioni2013}} and the more recent measurement by \citet{dsilva2023} that uses stellar population modelling of large galaxy redshift surveys to derive the star formation rate density.
The agreement between recovered CSFH curves using different SPLs is relatively good until 6 Gyr, beyond which the curves are more divergent, pointing to the underlying disagreements and uncertainties. 
This is consistent with the conclusions reached by \citet{bravo2022}, who determined that the recovery of galaxy properties from SED fitting was accurate to around 6 Gyr, beyond which modelling choices drove the recovered properties. 
It is notable that the CSFH derived using BC03, FSPS, and M05 all recover a substantial peak beyond 12 Gyr in lookback time, which is not reflected in the high-redshift  of \citet{madau2014} and \citet{dsilva2023} presented in the figure. 
A much smaller peak at this same epoch is produced by \textsc{ProGeny}, demonstrating that in four of the six SPLs studied in this work there is an over-assignment of ancient stars to galaxies. 
Furthermore, the observed peak of the CSFH at a lookback time of $\sim$10 Gyr is not well recovered by BC03, M05, CB19 and \textsc{ProGeny}, instead being much flatter between 6-12 Gyr. 
The overall shape of the CSFH is best recovered using FSPS, however the overall normalisation is slightly low (a consequence of the very high ancient peak). 

The remaining panels of Fig. \ref{fig:CSFH_all} present the derived CSFH using all tested SPLs with the alternative implementations explored in this work. 
From a visual comparison between the derived CSFH curves and the underlying observational measurements, it is clear that the fiducial implementation is better capable of recovering the observed CSFH. With the change to assuming a constant Z, it can be seen that the CSFH curves tend to start skewing toward younger peaks, which is a trend that continues further when moving to the $Z_{\odot}$ implementation in the third panel. It is also notable that there is a clear artefact recovered in the $Z_{\odot}$ implementation, where the SFRD values are overestimated around a Gyr in the past. The only SPL for which this behaviour is not present is M05.
While the overall recovery from the b5 SFH configuration in the bottom panel is similar to the fiducial, the impact of the stepwise parametrisation is clear in that the gradients are unable to be recovered. 
Recent work by \citet{mosleh2025} studying the recovery of SFHs in mock high-z galaxies showed that stepwise parametrisations were less able to recover bursts, and tended to overestimate SFRs in the earliest parts of the SFH. This is linked to the difficulty of these SFHs to recover the gradients of the SFHs. 
In the b5 SFH implementation it is noteworthy that the lowest-$z$ bin of the CSFH displays much more discrepancy between the SPL values than the other SED-fitting implementations. 
Note that the most recent burst of star formation available with the b5 SFH implementation hardly appears in this plot, as the CSFH only focuses on the region of lookback time that is covered by all galaxies, beyond $z=0.06$. This is an indication that the relative decoupling of the SFR values in neighbouring bins acts to reduce the constraint on slightly older stellar populations. 
In general, the quality of CSFH recovery by the b5 mode is similar to the recovery presented by \citet{leja2019a} using a stepwise parametric SFH with the code \textsc{Prospector}, where it was shown that the selection of a prior can also have a large impact on the SFH recovery. 

It is remarkable in Fig. \ref{fig:CSFH_all}, that as the underlying model becomes more physically motivated, so too does our capacity to infer realistic uncertainties. This is particularly clear when moving from the third, to the second, to the top panels, where the metallicity prescription becomes more realistic. While sampled ranges are similar in the last 6 Gyr, the uncertainties expand substantially for older times. 
The sampled uncertainties in the b5 SFH configuration (bottom panel) are almost negligible, indicating the relative lack of freedom that these broad bins provide.

\begin{figure}
	\centering
	\includegraphics[width=85mm]{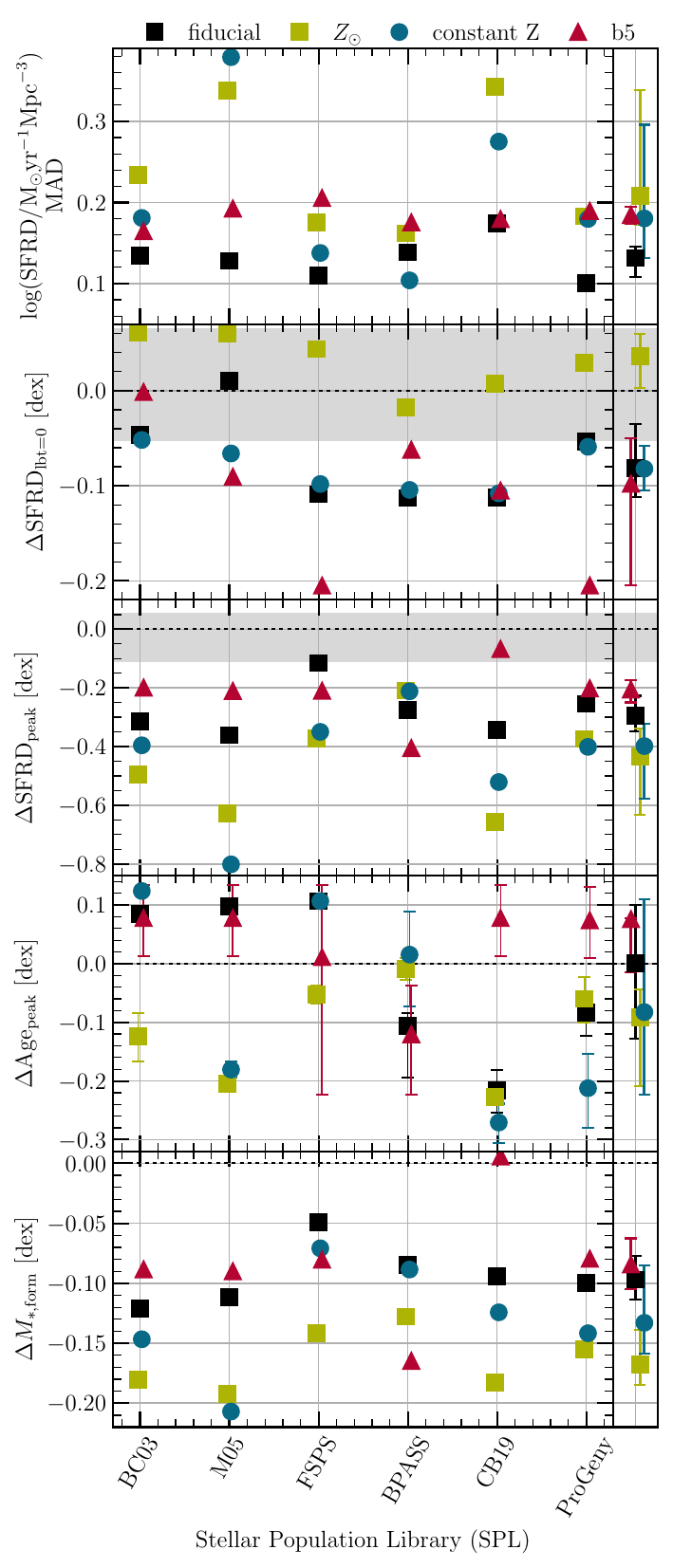}
	\caption{Numerical characterisation of the CSFH recovery by different SPLs, with varying SED fitting configurations. \textit{Top panel}: The mean absolute deviation from the CSFH for different stellar population libraries with different metallicity and SFH configurations. \textit{Second panel}: The difference in the recovered SFRD at the present day.  \textit{Third panel}: The difference in the recovered SFRD at the peak of the CSFH. \textit{Fourth panel}:The difference in age between the recovered peak and the observed peak (errorbars here convey the range covered by the peak, if extended). \textit{Bottom panel}: The difference in the integrated CSFH. The rightmost panel represents the median offset across all SPLs. }
	\label{fig:CSFH_meanAbsDeviation}
\end{figure}

A numerical comparison of the overall CSFH recovery is shown in Fig. \ref{fig:CSFH_meanAbsDeviation}. 
The top panel presents the mean absolute deviation (MAD) from the \citet{dsilva2023} CSFH for each SPL/configuration combination, computed over the lookback time range for which there are observations ($<$12.5 Gyr). The \textsc{ProGeny} fiducial run has the overall lowest MAD (only marginally better than FSPS). In general the fiducial configuration is best capable of replicating the CSFH, as indicated by the fact that the MAD averaged over SPLs (rightmost panel of Fig. \ref{fig:CSFH_meanAbsDeviation}) is the lowest for the fiducial implementation. 
BPASS is the only SPL for which the fiducial implementation does not produce the preferred CSFH. The different age--metallicity behaviour of BPASS as shown in Fig. \ref{fig:Age_Mass_Z} and discussed in Sec \ref{sec:AgeMetallicity} is deemed responsible for this, where the older ages are not preferred (likely due to the UV behaviour of high-metallicity templates) in the fiducial run when there is sufficient flexibility in metallicity.  

The second panel of Fig. \ref{fig:CSFH_meanAbsDeviation} presents the offset of the SFRD at lookback times of 0. 
The grey shaded region indicates the range covered by the observational uncertainty of \citet{dsilva2023}. 
In general, the $Z_{\odot}$ implementation can be seen to produce too much star formation, being the only configuration that over-predicts the current SFRD. However, this value is in best agreement with that of \citet{dsilva2023}. 
Conversely, the fiducial, constant Z, and b5 SFH implementations tend to produce underestimates. 
Overall, \textsc{M05}, BC03 and \textsc{ProGeny} produce a present-day SFRD that is most consistent with observations. This is perhaps unsurprising, as BC03 was used to infer the observational results by \citet{dsilva2023}. 

The third panel shows the offset of the SFRD at the observed peak of the CSFH. It is notable that the magnitude of the offsets at $\sim$10 Gyr in lookback time are substantially larger than present day values (by a factor of $\sim$6), demonstrating the increased difficulty of constraining star formation at older times. 
All SPLs underestimate the value of the peak, however FSPS is the closest to reproducing the observed SFRD value on average.
While the single implementation that gets closer to the SFRD peak is the CB19 b5 implementation, the other CB19 implementations (in particular those of the $Z_{\odot}$ and constant Z) have a much poorer reproduction of the SFRD peak. 
It is notable, that for the other SPLs that do not overall recover the peak star formation as well, the b5 SFH implementation instead produces the best peak recovery. 

The fourth panel shows the recovery of the age at which the peak in the CSFH occurs. On average the b5 SFH implementation produces the smallest offset, however the range is also very large, as the recovered SFRD is constant over around 6 Gyr with this method. The peak is frequently over- or under-estimated by up to 4 Gyr, showing the difficulty in recovering this property via SED fitting. 
Overall, FSPS and BPASS tend to be best performing in this space. 

Finally, the bottom panel shows the difference in the overall formed stellar mass density inferred between each SPL and the integrated CSFH from \citet{dsilva2023}. All implementations of SED fitting produce less mass than that inferred by observations, but FSPS is capable of producing the least offset values. 
In terms of the SED fitting implementation, the fiducial implementation with a snorm trunc SFH and evolving metallicity gets the closest to reproducing the overall amount of mass. We note that the CB19 b5 SFH implementation produces almost exactly the right amount of mass, however it seems likely that this is a case of the b5 SFH and CB19 biases cancelling out to produce a net better value, rather than being an indication that the b5 SFH or CB19 implementation produces better results in general.

\section{Impacts of SPL ingredients}
\label{sec:OtherImpacts}

In the previous section, we highlight that the choice in SPL produces a substantial impact on the resulting properties when conducting SED fitting on a sample of galaxies. 
In this section, we specifically isolate the impacts that individual ingredients into the SPL have on the CSFH, by changing individual ingredients within \textsc{ProGeny}, as the functionality exists here to do so. 
This expands on sec. 4 of \citet{robotham2024a}, which presents the impact of these individual ingredients on the \textsc{ProGeny} SPL and the SSPs generated. 
While the focus of this comparison is to assess the impact at the SED fitting stage, we highlight that a wealth of studies have been conducted to compare the outputs of simple stellar population models and the ingredients used to craft them on sources like globular clusters \citep[for example][]{charlot1996, maraston1998, thomas2003, maraston2005, percival2009, martins2019, coelho2020}.  

\subsection{Impact of the SPL Isochrone}

\begin{figure*}
	\centering
	\includegraphics[width=180mm]{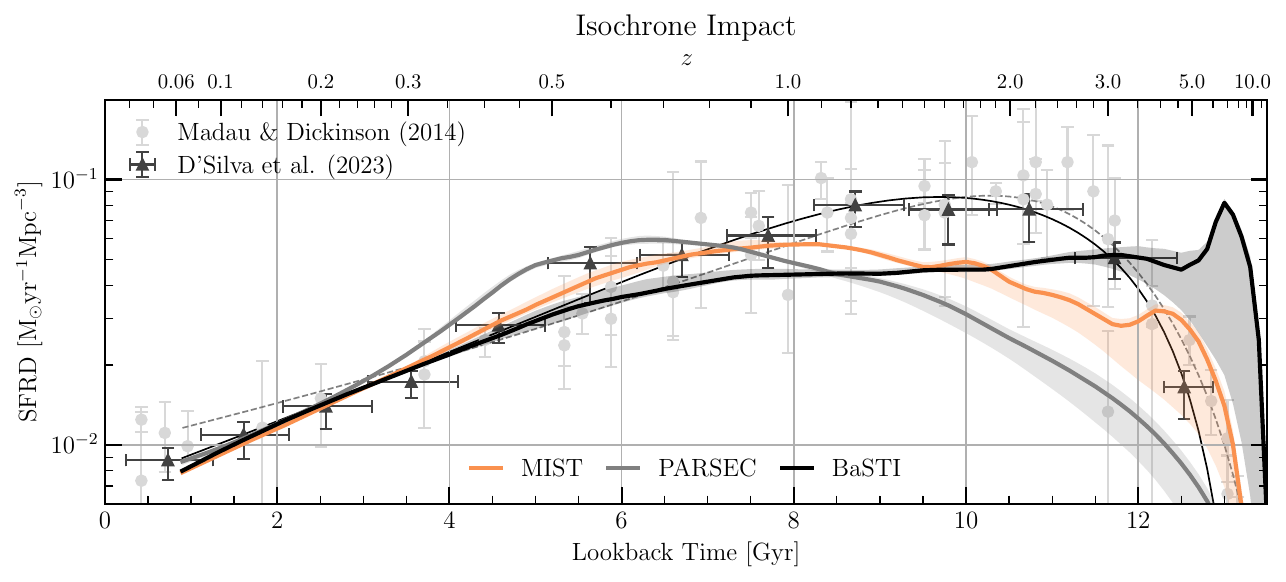}
	\caption{The CSFH as derived when using different isochrones within \textsc{ProGeny}, all assuming a Chabrier IMF and the C3K stellar spectra. }
	\label{fig:CSFH_isochrone}
\end{figure*}

\begin{figure}
	\centering
	\includegraphics[width=85mm]{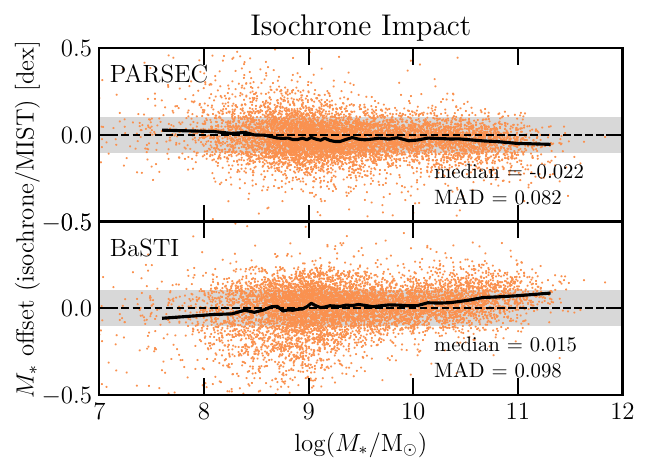}
	\caption{The stellar mass offets resulting when using different isochrones within \textsc{ProGeny}, all assuming a Chabrier IMF and the C3K stellar spectra. }
	\label{fig:Mass_isochrone}
\end{figure}

The flexibility of \textsc{ProGeny} allows us to demonstrate the impact of purely changing the isochrone origin of the SPL. Selecting the typical isochrones used in all the studied SPLs of this work. Fig. \ref{fig:CSFH_isochrone} shows the derived CSFH when using the MIST \citep{dotter2016} isochrones in orange (the default in the \textsc{ProGeny} outputs presented in this paper), as well as the PARSEC \citep{marigo2017} isochrones (very similar to the Padova isochrones, also used in BC03,  FSPS, and CB19, but including the COLIBRI tracks), and the BaSTI isochrones \citep[recently released, but as of yet not used in the other SPLs presented in this paper][]{hidalgo2018, pietrinferni2021, pietrinferni2024}. 
For consistency, a \citet{chabrier2003} IMF is assumed for all these outputs. 
PARSEC produces a CSFH with a peak skewed younger than the fiducial version using MIST, whilst the BaSTI isochrones result in a substantially older CSFH. 
The derived spread caused by these isochrones is generally consistent with the spread between different SPLs, indicating that a large fraction of the variation caused by the choice of SPL derived from the isochrone selection. 
The CSFH derived from all isochrones is exactly consistent within the last 2Gyr, showing that any discrepancies from isochrones are only expressed for older populations. 

Fig. \ref{fig:Mass_isochrone} presents the relative impact on the inferred stellar masses, highlighting that this is low on average (the maximum offset is $-0.022$ dex between PARSEC and MIST). We note though that while the median offset between BaSTI and MIST is low, it is mass-dependent, with the masses of high-mass galaxies predicted to be higher by BaSTI, but lower at lower stellar masses. 
In terms of the impact on SFR, the PARSEC-recovered SFRs have a median offset of  0.066 dex greater than MIST, while BaSTI and MIST produce much more consistent SFR values. 
On average both PARSEC and BaSTI produce younger half-mass ages than MIST (median offsets of $-0.035$ and $-0.023$ respectively). While for PARSEC this is consistent with the trend seen in the CSFH in Fig. \ref{fig:CSFH_isochrone}, the CSFH being skewed older when using the BaSTI isochrones is caused by the mass-dependent behaviour, where for galaxies with $\log(M_*/{\rm M}_{\odot}) > 10.5$ the half-mass ages are overestimated by around 0.05 dex. 

As presented by \citet{robotham2024a}, the MIST isochrones contain the largest population of models in the hot post-AGB\footnote{AGB - Asymptotic Giant Branch} phase, while BaSTI has a lower sampling, and PARSEC almost entirely neglects this stellar evolution phase. 
This is a suggestion that no post-AGB coverage (PARSEC) results in an age underestimate, whereas full post-AGB coverage (MIST) recovers older ages. The consequence of an only partial coverage (BaSTI) is a mass-dependent bias of the ages, where massive galaxies are overestimated in age, and low-mass galaxies are underestimated in age. 
Fig. \ref{fig:CSFH_isochrone} hence shows the impact of including this stellar evolution phase on SED fitting very clearly.

\subsection{Impact of the SPL Stellar Spectra}

\begin{figure*}
	\centering
	\includegraphics[width=180mm]{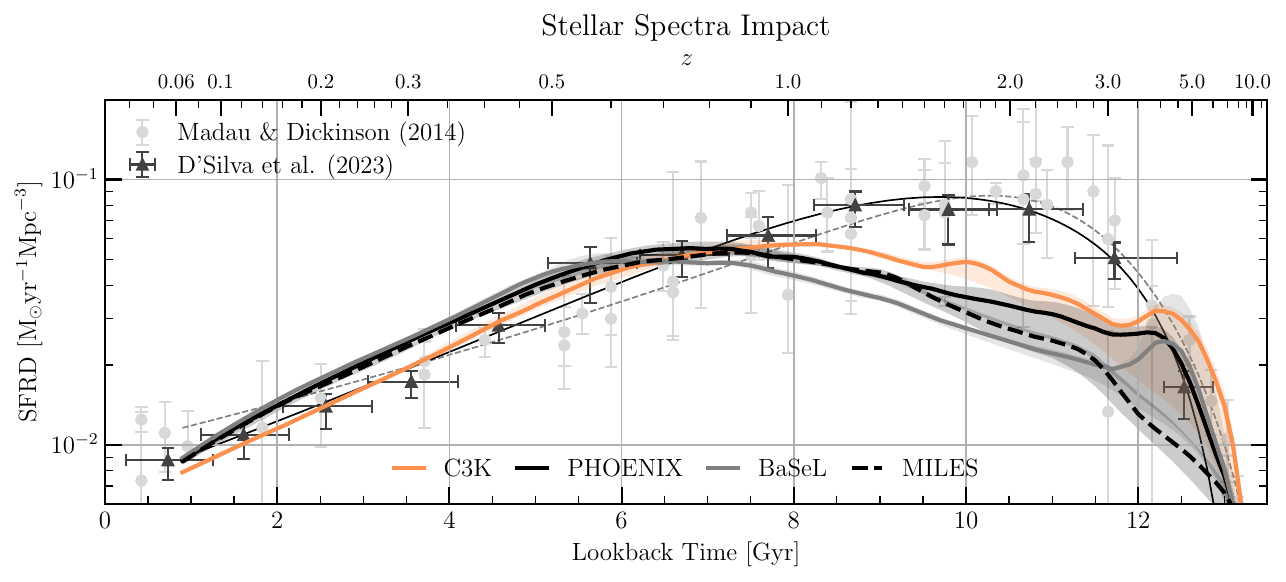}
	\caption{The CSFH as derived when using different stellar spectra within \textsc{ProGeny}, all assuming a Chabrier IMF and the MIST isochrones. }
	\label{fig:CSFH_spectra}
\end{figure*}

\begin{figure}
	\centering
	\includegraphics[width=85mm]{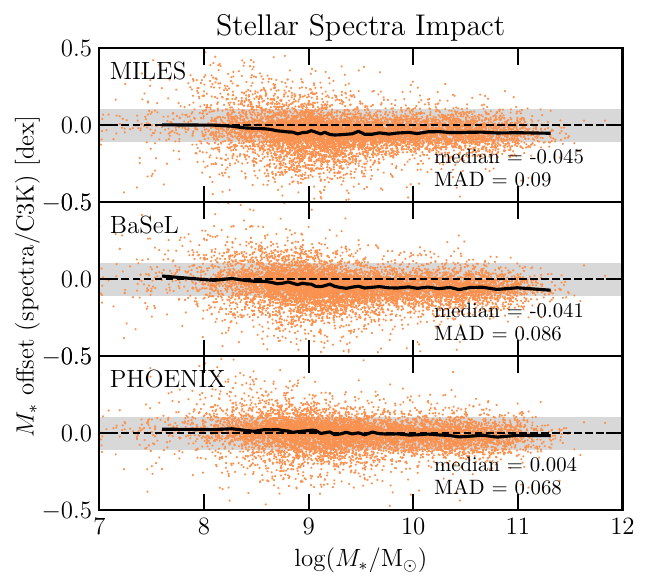}
	\caption{The stellar mass offets resulting when using different stellar spectra within \textsc{ProGeny}, all assuming a Chabrier IMF and the MIST isohrones. The fiducial version compared against uses the C3K stellar spectra.  }
	\label{fig:Mass_spectra}
\end{figure}

The stellar spectra used in any given SPL determine the selected spectrum used for any stellar property specified by the isochrones. 

As described in detail by \citet{robotham2024a}, \textsc{ProGeny} has been constructed in such a way as to combine multiple sets of astmospheres to facilitate maximal coverage of the isochrones. 
This is facilitated by setting a `base' set of templates to cover most stars, which can be supplemented by an `extend' set from a different set of stellar spectra to improve coverage. 
\textsc{ProGeny} supports various different publicly-available stellar spectra sets, including C3K \citep{conroy2018}, MILES \citep{sanchez-blazquez2006}, PHOENIX \citep[two versions of which are available][]{allard2012, husser2013}, and BaSeL \citep{lejeune1997}. 
A summary of the properties defining these stellar spectra sets is provided in table 3 of \citet{robotham2024a}. 
C3K and PHOENIX are both synthetically generated (C3K derived via the ATLAS and SYNTHE codes \citep{kurucz1993}, and PHOENIX is derived using the code of the same name \citep{hauschildt1999}), while MILES is empirical, and BaSeL semi-empirical. 
MILES is supported by FSPS, and the version used within \textsc{ProGeny} is taken directly from FSPS. Of particular note, is that the property gridding is adapted to be consistent with BaSeL, and furthermore the wavelength range is extended to that used in BaSeL as well (different from the wavelength extension implemented in EMILES \citealt{vazdekis2016}). 
BaSeL is used broadly across different SPLs, including BC03, BPASS, M05, CB19 and FSPS. 
In the fiducial implementation of \textsc{ProGeny} in this paper, the `base' stellar spectra used are from C3K, with the `extend` set selected from the \citet{allard2012} version of the PHOENIX spectra. 
Using the MIST isochrones and Chabrier IMF fixed for each implementation, we vary the input `base' stellar spectra to quantify the impact of these stellar spectra sets on the derived properties from SED fitting. 
When modifying the stellar spectra being used, it is the `base' being modified, and in all cases the \citet{allard2012} version of the PHOENIX spectra are used as the supplemented `extend' set. 
For our PHOENIX version, the base set is selected as the \citet{husser2013} version of the PHOENIX spectra. 

As in Fig. \ref{fig:CSFH_isochrone}, Fig. \ref{fig:CSFH_spectra} shows the impact on the resulting CSFH that purely changing the stellar spectra sets causes when used in SED fitting. 
The impact on the CSFH is similar to the impact of a constant metallicity assumption seen in the SED fitting configurations. 
Certainly the variation is greater than that recovered through sampling uncertainty in the SED fitting. 
The fiducial C3K stellar spectra recover the oldest CSFH, while PHOENIX, BaSeL, and MILES all recover a CSFH that slightly overestimates the SFRD below 7 Gyr in lookback time. 
The present-day SFRD value recovered using C3K is slightly lower than the value recovered with all other stellar spectra sets, however all still within the uncertainty of the \citet{dsilva2023} observational measurement. 
Fig. \ref{fig:Mass_IMF} shows the resulting offsets in stellar masses, which are seen to be very low (the biggest discrepancy is between C3K and MILES, producing a median offset of 0.045 dex). 
Correspondingly, the resulting offsets in SFR are low (a maximum offset of 0.031 dex between C3K and MILES, where MILES recovers larger SFRs on average), and the maximum offset in age is between C3K and MILES, at 0.051 dex (consistent with the trend seen in the CSFH in Fig. \ref{fig:CSFH_spectra}). 

Identifying the specific origin for these discrepancies is difficult, due to the large number of steps between spectrum selection and the final galaxy properties being derived. In each of these steps, a multitiude of small differences can have cascading effects. C3K produces a CSFH that is most unique (whereas PHOENIX, BaSeL, and MILES have CSFHs that are more similar, in particular within the last 7 Gyr), however based on the main properties of the stellar spectra sets (wavelength range, theoretical/observational origin, metallicity range etc) there is nothing that particularly distinguishes C3K.

\subsection{Impact of the SPL Initial Mass Function}

\begin{figure}
	\centering
	\includegraphics[width=85mm]{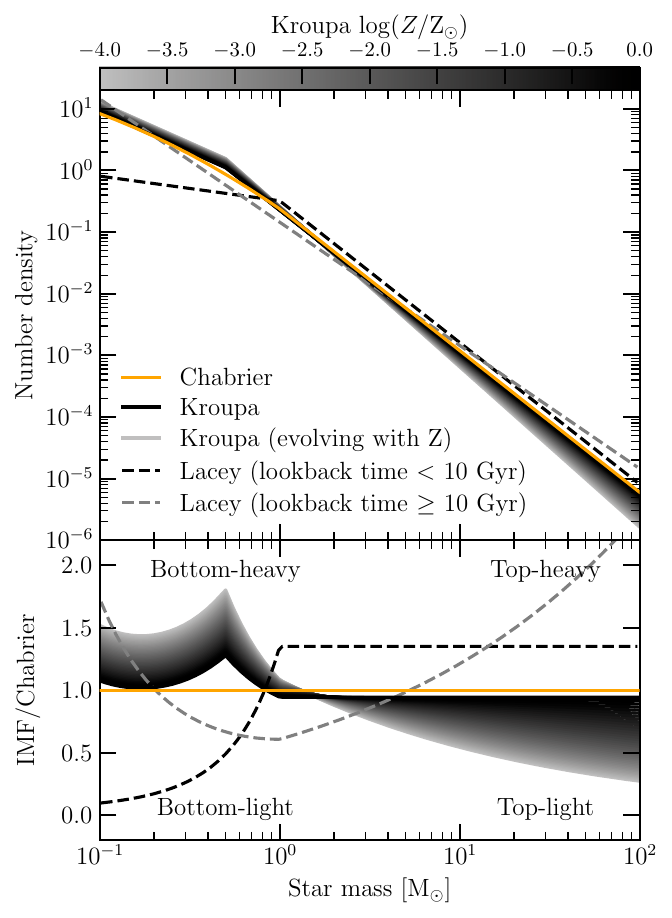}
	\caption{The Initial Mass Functions (IMFs) tested in this work (top panel), and their residual relative to the fiducial Chabrier IMF (bottom panel). All IMFs have been normalised to show the same amount of stellar mass. }
	\label{fig:IMF}
\end{figure}

\begin{figure*}
	\centering
	\includegraphics[width=180mm]{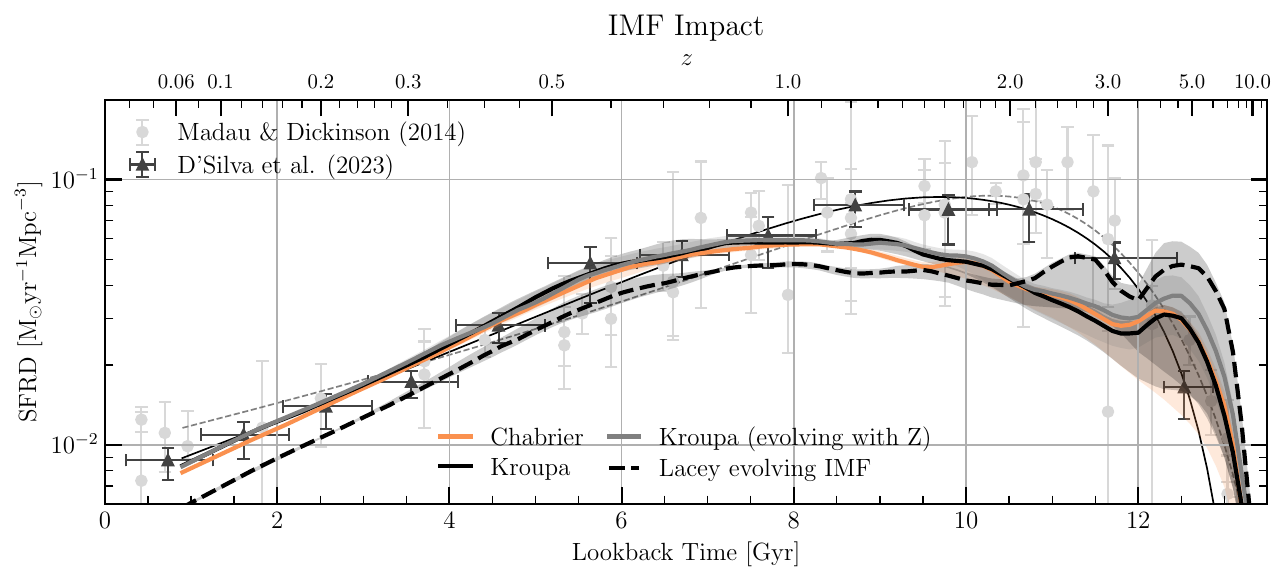}
	\caption{The CSFH as derived when using different IMF prescriptions within \textsc{ProGeny}. }
	\label{fig:CSFH_IMF}
\end{figure*}

\begin{figure}
	\centering
	\includegraphics[width=85mm]{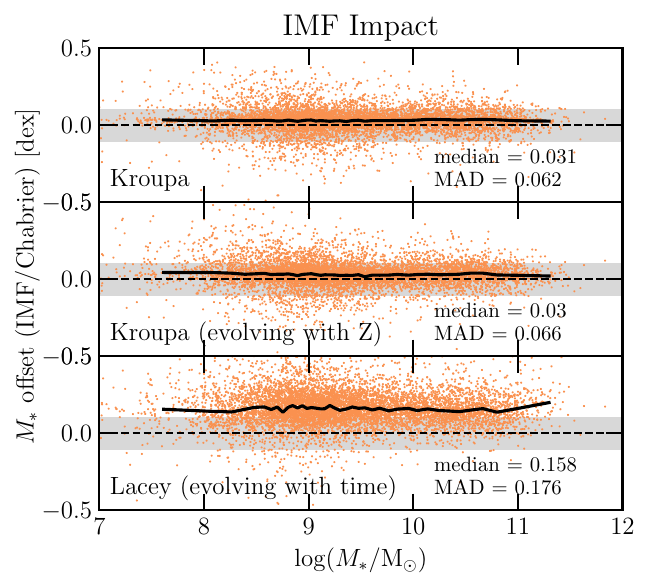}
	\caption{The stellar mass offets resulting when using different IMF prescriptions within \textsc{ProGeny}. }
	\label{fig:Mass_IMF}
\end{figure}

Making further use of the flexibility of \textsc{ProGeny}, we conduct a number of tests to isolate the impact of different Initial Mass Functions (IMFs)\footnote{The code to produce the \textsc{ProGeny} SPLs using each of the implemented IMFs is available online at \url{https://rpubs.com/asgr/1235271}.} on the SED fitting output. 
We note that while \citet{robotham2024a} made comparisons against a \citet{salpeter1955} IMF, we do not include this in our comparisons, as it is generally regarded to be too bottom heavy for low-$z$ galaxy SED fitting. 
For consistency, all IMFs are constructed using lower and upper mass limits of 0.1 $\rm{M}_{\odot}$ and 100 $\rm{M}_{\odot}$ respectively.  
We plot all IMFs in Fig. \ref{fig:IMF}, where the IMF itself is shown in the top panel, and the residual to the fiducial Chabrier IMF in the bottom panel. In the bottom panel we highlight the four regions of parameter space where the terms `top-heavy', `top-light', `bottom-heavy', and `bottom-light' are used to describe the shape of th IMF relative to Chabrier. 

Fig. \ref{fig:CSFH_IMF} shows the derived CSFH of the fiducial run using \textsc{ProGeny} with a \citet{chabrier2003} IMF, as well as the equivalent when assuming a \citet{kroupa2002} IMF. 
It is notable that the variation between these two is exceptionally small - smaller even than the impact of assuming the metallicity to be constant or evolving. The main discernable difference between a Chabrier and Kroupa IMF in terms of the recovered CSFH, is the slightly higher inferred SFRs in the last $\sim$ 2 Gyr when a Kroupa IMF is assumed (although this difference is well within the uncertainties of the observed CSFH). 
The corresponding impact on the derived stellar masses is shown in the top panel of Fig. \ref{fig:Mass_IMF}. 
Because the Kroupa IMF has slightly more low-mass stars than the Chabrier IMF, it is unsurprising that the stellar msses derived using the Kroupa IMF are systematically higher by around 0.03 dex. 
Similarly, we find that the estimated SFR values are also higher using Kroupa by around 0.03 dex. 

To assess the impact of a more ``exotic" treatment of the IMF, we also produce a simple stellar population using \textsc{ProGeny} that evolves in time according to the \citet{lacey2016} prescription. 
In this prescription, a simple temporal evolution is provided through a star burst dominated top-heavy IMF at high redshift (used by \citealt{lacey2016} for the ``starburst" mode, presented as the dashed grey line in Fig. \ref{fig:IMF}), transitioning to a \citet{kennicutt1983} IMF at a lookback time of 10 Gyr (used by \citealt{lacey2016} for the ``quiescent" mode, presented s the dashed black line in Fig. \ref{fig:IMF}). Note from Fig. \ref{fig:IMF} that while most top-heavy IMFs are bottom-light (and vice versa), the starburst mode from \citet{lacey2016} occupies both the top-heavy region and the bottom-heavy, after transitioning through the bottom-light quadrant. 
Both the starburst and quiescent mode IMF prescriptions are more top-heavy than the \citeauthor{kroupa2002} IMF, with differences at the low-mass end toward a more bottom-heavy IMF for the starburst mode and more bottom-light for the quiescent mode. 

The impact of this temporally evolving IMF is seen to be greater than simply the swap from a \citeauthor{chabrier2003} to a \citeauthor{kroupa2002} IMF (particularly in the last 5 Gyr, where the evolving IMF is seen to systematically recover lower SFRs). 
Of note as well, is that while the transition of IMFs at a lookback time of 10 Gyr in the Lacey prescription is somewhat arbitrary and abrupt (selected to coincide with cosmic noon), there is no corresponding impact of this seen in the CSFH, in terms of a discontinuity. 
This is however the transition point between two epochs that demonstrate different behaviour. 
In the older epoch Lacey-derived CSFH has a greater SFRD than the Chabrier-derived one, while in the younger epoch the Lacey CSFH has a lower SFRD than the Chabrier one. 
The impact on the derived stellar masses is seen to be substantial, in the bottom panel of Fig. \ref{fig:Mass_IMF}, with a systematic increase in 0.158 dex. 
As is evident from the lower SFRD at present day in Fig. \ref{fig:CSFH_IMF}, the corresponding SFRs are also recovered to be systematically 0.134 dex lower than with the assumption of a Chabrier IMF. 

Finally, we also test an implementation of an evolving IMF where the IMF changes as a function of metallicity, rather than a function of time \citep[motivated by resolved galaxy observations such as those by][who measure a dependence of the IMF on the local metallicity]{martin-navarro2015}. 
In this implementation, the Kroupa IMF slope in the [0.5$\rm{M}_{\odot}$, 150$\rm{M}_{\odot}$] mass range ($\alpha_3$) varies between 2 to 2.3 depending on the metallicty, mapped linearly from the metallicity range -4 to 0, in units of $\log(Z/Z_{\odot})$. 
The implication of this implementation is that at higher metallicities the IMF is exactly Kroupa, but at lower metallicities the high-mass slope becomes steeper (thereby becoming more top-light). 
The slopes at the lower mass ranges ($\alpha_1$ and $\alpha_2$) remain fixed at the values 0.3 and 1.3 respectively.  
We highlight that these values were only roughly chosen as a demonstration of how a metallicity-dependent IMF impacts SED fitting. 
Further work is required to refine such an implementation for future applications, and we point the reader to very recent work by \citet{bate2025} who show using simulations that the IMF likely varies with both metallicity and redshift, becoming more bottom-light with both increasing metallicity and redshift. 
The resulting CSFH from this IMF implementation is shown in Fig. \ref{fig:CSFH_IMF} by the dashed black line. It is noteworthy that the impact of this evolution is only very subtle, with the effect of slightly increasing the SFRD at early times. 
The middle panel of Fig. \ref{fig:Mass_IMF} shows that again the impact on stellar masses is systematic, with a median offset of 0.03 dex. We find that the overall median impact on SFRs is 0.001, however this is stellar-mass dependent, with low-mass galaxies resulting in a slight SFR underestimate using the evolving Z IMF and high-mass galaxies resulting in a slight SFR overestimate using the evolving Z IMF.

\section{Discussion}
\label{sec:Discussion}

\subsection{CSFH impact from SPL sub-ingredients}

\begin{figure}
	\centering
	\includegraphics[width=85mm]{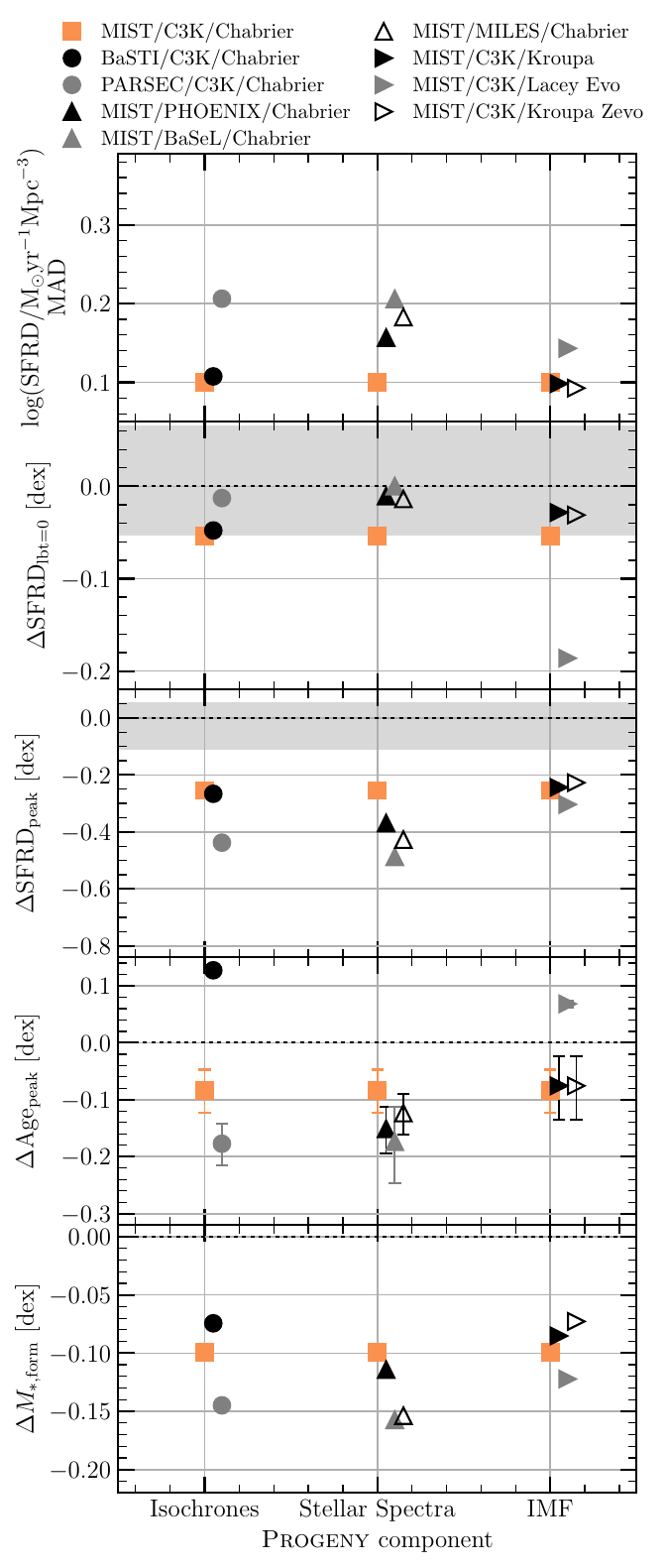}
	\caption{Equivalent of Fig. \ref{fig:CSFH_meanAbsDeviation} to characterise the impact on the derived CSFH when modifying only sub-ingredients of the \textsc{ProGeny} SPL. The orange points represent the fiducial implementation of \textsc{ProGeny} using the MIST isochrones, the C3K stellar spectra, and a Chabrier IMF (identical to the black \textsc{ProGeny} point plotted in Fig. \ref{fig:CSFH_meanAbsDeviation}). Each other point point to a result after a single modification in the \textsc{ProGeny} SPL.  }
	\label{fig:CSFH_meanAbsDeviation_SPLingredients}
\end{figure}

We present the equivalent of Fig. \ref{fig:CSFH_meanAbsDeviation} for only the impact of modifying the sub-ingredients of the \textsc{ProGeny} SPL in Fig. \ref{fig:CSFH_meanAbsDeviation_SPLingredients}. The top panel of Fig. \ref{fig:CSFH_meanAbsDeviation_SPLingredients} shows that the only modification of the \textsc{ProGeny} SPL that results in an improved recovery of the CSFH is a marginal improvement seen when changing the IMF to either a Kroupa IMF, or a Kroupa IMF evolving with the metallicity. 

The second panel shows that other modifications result in a SFRD at the present day more similar to the value from \citet{dsilva2023}, however all except the Lacey IMF modification produce a value consistent within observational uncertainty. The SFRD at the CSFH peak (third panel) and the age of the peak (shown in the fourth panel) is marginally better recovered by the Kroupa and evolving Kroupa modifications than the fiducial implementation. The total amount of mass formed in each modification relative to the \citet{dsilva2023} CSFH is then shown in the bottom panel. 

\subsection{Ranking of modelling assumptions}

Any single change in the implementation of SED fitting will impact the outputs. There have been many studies that have categorised these impacts individually \citep[for example][]{carnall2019, leja2019a, lower2020, suess2022, thorne2022, jones2022, haskell2024}, and the results of this study show that some impacts are sub-dominant to others. 

The impact of stellar library ingredients and SED fitting implementations on the derived stellar masses, SFRs, and half-mass ages of the overall population are summarised in Tab. \ref{tab:MassImpact}. 
Potential configuration modifications have been ranked from those that have the largest impact, to the smallest impact in terms of the largest median offset.

\begin{table*}
	\centering
	\caption[MassImpact]{The relative difference range in derived stellar masses, SFRs, and half-mass age values when changing specific aspects of the SED fitting. For each implementation modification, the range in total median offsets is provided, as well as the range in MAD values. Modifications are ranked from greatest to least impact on the derived mass.  }
	\label{tab:MassImpact}
	\begin{tabular}{@{}c | c c p{0.1cm}  | c c p{0.1cm}| c c p{0.1cm}  | p{0.2cm}}
		\hline
		Modification & \multicolumn{3}{c}{Stellar Mass Offset (dex)}& \multicolumn{3}{c}{SFR Offset (dex)}&\multicolumn{3}{c}{Half-Mass Age Offset (dex)}  & Rank\\
		&  Offset & MAD & \#& Offset  & MAD & \#& Offset &MAD & \# & \\
		\hline
		\hline
		$Z_{\odot}^{\dagger}$ & [$-0.27$, $-0.12$] &  [$0.18$, $0.31$] &1 & [$0.026$, $0.20$]&  [$0.46$, $0.82$]  & 1& [$-0.47$, $-0.35$] &  [$0.40$, $0.49$] &1  & 1\\
		\hline
		b5 SFH$^{\dagger}$  & [$0.048$, $0.11$] & [$0.09$, $0.17$]& 4 & [$-0.00018$, $0.11$] &  [$0.47$, $0.82$] & 3& [$0.06$, $0.12$]&  [$0.15$, $0.23$]  & 2& 2 \\
		\hline
		SPL$^{\dagger}$ & [$-0.15$, $0.027$] & [$0.083$, $0.18$] &3 & [$-0.087$, $0.092$] &  [$0.097$, $0.18$] & 4 & [$-0.081$, $0.022$] &  [$0.13$, $0.20$] &3 & 3\\
		\hline
		IMF$^{\ddagger}$ & [$0.03$, $0.16$] & [$0.062$, $0.18$]& 2 & [$-0.13$, $0.027$]&  [$0.16$, $1.8$]  &2 & [$-0.006$, $0.016$] &  [$0.093$, $0.11$] &7  & 4 \\
		\hline
		Stellar Spectra$^{\ddagger}$ & [$-0.045$, $0.004$] & [$0.068$, $0.09$]& 5 & [$0.021$, $0.031$]&  [$0.099$, $0.102$]  &5 & [$-0.051$, $-0.019$] &  [$0.11$, $0.13$] & 4 & 5\\
		\hline
		Isochrone$^{\ddagger}$ & [$-0.018, 0.016$] & [$0.082$, $0.098$] &6  & [$-0.006$, $0.068$]&  [$0.11$, $0.13$]  & 7& [$-0.035$, $-0.023$] &  [$0.12$, $0.15$] &5 & 6 \\
		\hline
		Constant Z$^{\dagger}$ & [$-0.010, 0.004$] & [$0.053$, $0.11$]& 7 & [$-0.027$, $-0.004$] &  [$0.35$, $0.63$] & 6& [$-0.018$, $0.0$]&  [$0.09$, $0.13$]  & 6 & 7\\
		\hline		
	\end{tabular}
	\begin{tablenotes}
		\small
		\item$\dagger$ Comparison is conducted by implementing configuration for six SPLs (BC03, M05, FSPS, BPASS, CB19, and fiducial \textsc{ProGeny}). 
		\item$ \ddagger$ Comparison is conducted by modifying the option within \textsc{ProGeny}. 
	\end{tablenotes}
\end{table*}

If retrieving accurate stellar masses is the primary aim of SED fitting, then the greatest bias will result from fixing the modelled metallicity to solar. 
While the impact of changing IMF from Chabrier to Kroupa is relatively small, the potential impact from assuming a more exotic IMF evolution is high (0.16 dex). 
SPL selection is the next most impactful choice, with a change to BPASS causing the greatest offset of $-0.15$ dex. 
The change of SFH parametrisation to b5 SFH creates a maximum median offset of 0.11 dex (consistent with similar estimates recorded in the literature). 
The specific modifications within an SPL (specifically the changing of stellar spectra and isochrones) have lower impacts, with maximum offsets of -0.045 and -0.018 respectively. 
Finally, the change to modelling a metallicity history as constant versus evolving has a negligible impact on stellar masses. 

If instead retrieving the SFRs is the primary aim of SED fitting, then we show that again the greatest bias is caused by fixing the modelled metallicity to solar. 
The bias is systematically positive, with the smallest offset identified when using the CB19 SPL, and the greatest offset found with M05 (followed closely by BPASS). 
While the impact of a Kroupa versus Chabrier IMF on the SFRs is small, an exotic form of IMF evolution results in the second-largest impact on SFRs after the fixed $Z_{\odot}$ assumption. 
These assumptions are followed by the b5 SFH parametrisation. While an offset of up to 0.11 dex can be recovered through use of b5 SFH, the large MADs indicate that the scatter caused by the SFH parametrisation is also large (also evident from Fig. \ref{fig:SFROffset}). 
SPL choice has a noticeable impact on SFRs, and then the impacts of stellar spectra, isochrones, and constant Z are much smaller.  

While the half-mass age does not capture the complexities of star formation histories, Table \ref{tab:MassImpact} also presents the ranked list of assumptions that impact the recovered ages. This ranked list differs slightly from the other global parameters. 
Once again, the fixed metallicity assumption has the largest impact, underestimating ages by up to 0.47 dex.
The impact of b5 SFH is the second largest, with an up to 0.12 dex overestimate recorded for BPASS, CB19 and ProGeny. 
The jump down in biases is smaller for the SPL, which can cause a variation of 0.1 dex in age values. Stellar spectra and isochrones cause a similarly small offset, followed by the constant Z assumption, and finally the variation in IMF can be seen to have a minor impact on the recovered half-mass age.

While Table \ref{tab:MassImpact} ranks the factors that impact the recovery of global galaxy parameters, we rank the factors separately that impact the ability to recover the CSFH in Table \ref{tab:CSFHImpact_relativeToFiducial}. In contrast to the MAD values presented in Fig. \ref{fig:CSFH_meanAbsDeviation}, this table demonstrates the deviation between the base CSFH and the modified version, as a mechanism of conveying relative impact. It is notable that this ranked list is different, with the constant Z assumption driving one of the largest deviations from the CSFH (where M05 is most discrepant), and the median impact of stellar spectra and isochrones within the SPL creating a similar impact to that of the constant Z modification, followed closely by the b5 SFH. By far the smallest impact on the CSFH recovery is caused by the IMF selection. 


\begin{table}
	\centering
	\caption[MassImpact]{The relative difference range in CSFH Mean Absolute Deviations (MAD), when compared against the fiducial CSFH. Modifications are ranked from greatest to least median impact on the derived CSFH.  }
	\label{tab:CSFHImpact_relativeToFiducial}
	\begin{tabular}{@{}c | c c | c}
		\hline
		Modification & \multicolumn{2}{c}{CSFH MAD} & Rank\\
		& Range & Median & \\
		\hline
		\hline
		$Z_{\odot}$$^{*}$ & [$0.11$, $0.24$] & 0.17 &1 \\
		\hline
		Constant Z$^{*}$ & [$0.07$, $0.29$] & 0.10  & 2\\
		\hline
		Stellar Spectra$^{\ddagger}$ & [$0.07$, $0.12$] & 0.10 & 3 \\
		\hline
		Isochrone$^{\ddagger}$ & [$0.07$, $0.12$] & 0.10 & 3 \\
		\hline		
		b5 SFH$^{*}$  & [$0.09$, $0.25$] & 0.09 &5 \\
		\hline
		SPL$^{\dagger}$ & [$0.06$, $0.13$] & 0.08  &6\\
		\hline
		IMF$^{\ddagger}$ & [$0.02$, $0.09$] &  0.02 &7 \\
		\hline
	\end{tabular}
	\begin{tablenotes}
		\small
		\item$*$ Comparison is conducted by modifying the configuration for each given SPL. 
		\item$\dagger$ Comparison is conducted by implementing configuration for six SPLs (BC03, M05, FSPS, BPASS, CB19, and fiducial \textsc{ProGeny}). 
		\item$\ddagger$ Comparison is conducted by modifying the option within \textsc{ProGeny}. 
	\end{tablenotes}
\end{table}

We aim for these ranked lists to be useful for not only providing guidance in terms of the SED fitting aspects that should be prioritised for accurate results, but also to provide direction as far as the most important areas where improvement is still required. 
For example, to most accurately infer star formation histories in galaxy populations, it is unlikely that understanding the nuances of IMF evolution will improve our results, rather is it more critical for our most used isochrones to capture the full extent of stellar evolution. 
While the values in Tab. \ref{tab:MassImpact} are certainly useful in quantifying the property uncertainty introduced through specific choices in SED fitting and SPL generation, we strongly encourage the reader not to use these values as corrections.

\subsection{Modelling assumptions not considered in this work}

While we have compared a wide array of different inputs and modelling assumptions to assess the impact on derived SED fitting outputs, there still remain a number of SED fitting aspects that we have not considered in this work. 

Most notably, we do not assess the impact of dust prescriptions \citep[as done by][]{jones2022}, and we have employed a uniform dust prescription in all fits, where only the opacity and temperatures ($\tau$ and $\alpha$ parameters) of the birth cloud and ISM are allowed to be free. This does not include any prescriptions of the dust emission beyond that provided by \citet{dale2014}, nor does it consider any prescription of the dust absorption beyond \citet{charlot2000}. 

Another aspect in SED fitting that will impact the ability to recover realistic properties, is the inclusion of burstiness in a SFH. Recent work has highlighted the potential impact of burstiness \citep[for example][]{iyer2019, iyer2024, sun2023, narayanan2025, haskell2024}, however the difficulty in fitting burstiness for individual galaxies means that for now this is limited to studying averaged behaviour over populations.

We have not considered the impact of AGN at any stage of this work, both in terms of AGN contamination in the sample, or from the impact of fitting AGN as an additional component in the SED fitting. While we are confident that the AGN contamination rate in the sample is low (meaning that we are unlikely to be mis-modelling AGN flux as a stellar population), the relative biases measured in different configurations may be different in a sample where AGN are prevalent, and being fitted by additional components (in a mode more similar to that conducted by \citealt{thorne2022}). Therefore we highlight that the biases and impacts studied in this work really are limited to the stellar component of SED fitting. 

Finally, this study has focussed solely on a sample of low-redshift galaxies. 
When SED-fitting techniques are applied to high-$z$ galaxies, then the total amount of cosmic time requiring modelling is substantially reduced, thereby changing the nature of the systematics at play. High-redshift fitting may therefore present a much more favourable application for stepwise parametric SFHs, where individual bins are not required to cover such large time periods. Indeed, much recent SFH fitting of high-$z$ galaxies has been conducted in this way, as for example in \citet{nanayakkara2025}, however functional parametric forms are still proving popular, as for example in \citet{carnall2023}.  Furthermore, the nature of star formation is much more extreme in the high-redshift Universe, and it is quite likely that dust properties vary \citep[see for example][]{shivaei2024}. 

\section{Conclusions}
\label{sec:Conclusions}

We have conducted an SED-fitting analysis using the code \textsc{ProSpect} on a volume-complete sample of 7,862 galaxies from the GAMA survey, to characterise the impact of SPL selection against the impact of SED-fitting modelling choices like the metallicity implementation and SFH parametrisation. Furthermore, we quantify the impact of sub-SPL changes, including isochrone selection, stellar spectra selection, and the chosen IMF implementation. 
A number of conclusions have been reached from this analysis:

\begin{itemize}
	\item The greatest bias we find in estimating galaxy properties from SED fitting is caused by fixing the galaxy metallicity to $Z_{\odot}$, which has the potential to underestimate stellar masses by 0.27 dex, overestimate SFRs by 0.2 dex, and underestimate half-mass ages by 0.47 dex. Furthermore, this assumption typically produces the greatest impact on the inferred cosmic star formation history (CSFH). 
	\item The biases introduced through the choice of stellar poulation library (SPL) in SED fitting are of a similar magnitude to those introduced by choices like the SFH parametrisation, in terms of the derived properties of stellar mass, SFR, and half-mass age. While some contributing factors relate to underlying physics that changes (such as a different IMF or method of deriving stellar isochrones), other aspects of the implementation also have substantial impact that are not rooted in physics (such as the interpolation mechanisms used). 
	\item The evolving metallicity prescription is found to have relatively little impact on global galaxy properties like stellar mass and SFR, however it removes systematic biases in the derived CSFH and produces more realistic uncertainties in the recovered SFHs. To recover the observed CSFH, it is essential to include an evolving metallicity. 
	\item Sampled uncertainties in the fitting of SFHs vastly underestimate the true uncertainties introduced through SPL selection and modelling choices. 
	\item Both the isochrone selection and the stellar spectra selection are significant factors in influencing the differences in recovered SFHs at the SED fitting level, each with a potential impact equivalent to the constant metallicity assumption and the change in SFH parametrisation. 
	\item The impact of IMF is very small as compared to all other effects studied in this work in terms of the recovered SFHs. However, the selection of IMF is important in terms of estimating stellar masses and SFRs. 
\end{itemize}

\section*{Data Availability}

The data used as an input to this analysis are all publicly available via the GAMA webpage\footnote{\url{http://www.gama-survey.org}}. 
All software used for this analysis (the SED fitting code \textsc{ProSpect} and the SPL generating code \textsc{ProGeny}) is publicly available via Github\footnote{\url{https://github.com/asgr}}.
The modelling output presented in this paper can be made available upon reasonable request.

\section*{Acknowledgements}

The authors thank the anonymous referee for their detailed reading of the paper, which has resulted in multiple improvements. 
The authors also acknowledge the contributions made by Luke Davies and Simon Driver through various discussions and the anonymous referee via their helpful recommendations. 
ASGR and SB acknowledge support from the ARC Future Fellowship scheme (FT200100375). 
SB acknowledges funding by the Australian Research Council (ARC) Laureate Fellowship scheme (FL220100191).
This work was supported by resources provided by the \textit{Pawsey Supercomputing Centre} with funding from the Australian Government and the Government of Western Australia.
In total, around 174,000 CPU hours were required for the testing and final computation of all outputs presented in this work. 

We have used \textsc{R} \citep{rcoreteam2017} and \textsc{python} for our data analysis, and acknowledge the use of \textsc{Matplotlib} \citep{hunter2007} and \textsc{CMasher} \citep{vandervelden2020} for the generation of plots in this paper. This research made use of \textsc{Pandas} \citep{mckinney2010}, and \textsc{NumPy} \citep{harris2020}.

\bibliographystyle{mnras}
\setlength{\bibsep}{0.0pt}
\bibliography{ZoteroLibrary}

\begin{thebibliography}{}
\makeatletter
\relax
\def\mn@urlcharsother{\let\do\@makeother \do\$\do\&\do\#\do\^\do\_\do\%\do\~}
\def\mn@doi{\begingroup\mn@urlcharsother \@ifnextchar [ {\mn@doi@}
  {\mn@doi@[]}}
\def\mn@doi@[#1]#2{\def\@tempa{#1}\ifx\@tempa\@empty \href
  {http://dx.doi.org/#2} {doi:#2}\else \href {http://dx.doi.org/#2} {#1}\fi
  \endgroup}
\def\mn@eprint#1#2{\mn@eprint@#1:#2::\@nil}
\def\mn@eprint@arXiv#1{\href {http://arxiv.org/abs/#1} {{\tt arXiv:#1}}}
\def\mn@eprint@dblp#1{\href {http://dblp.uni-trier.de/rec/bibtex/#1.xml}
  {dblp:#1}}
\def\mn@eprint@#1:#2:#3:#4\@nil{\def\@tempa {#1}\def\@tempb {#2}\def\@tempc
  {#3}\ifx \@tempc \@empty \let \@tempc \@tempb \let \@tempb \@tempa \fi \ifx
  \@tempb \@empty \def\@tempb {arXiv}\fi \@ifundefined
  {mn@eprint@\@tempb}{\@tempb:\@tempc}{\expandafter \expandafter \csname
  mn@eprint@\@tempb\endcsname \expandafter{\@tempc}}}

\bibitem[\protect\citeauthoryear{Allard, Homeier  \& Freytag}{Allard
  et~al.}{2012}]{allard2012}
Allard F.,  Homeier D.,   Freytag B.,  2012, \mn@doi [Philosophical
  Transactions of the Royal Society A: Mathematical, Physical and Engineering
  Sciences] {10.1098/rsta.2011.0269}, 370, 2765

\bibitem[\protect\citeauthoryear{Bate}{Bate}{2025}]{bate2025}
Bate M.~R.,  2025, Variation of the Low-Mass End of the Stellar Initial Mass
  Function with Redshift and Metallicity (\mn@eprint {arXiv} {2501.06082}),
  \mn@doi{10.1093/mnras/staf059}

\bibitem[\protect\citeauthoryear{Bellstedt et~al.,}{Bellstedt
  et~al.}{2020a}]{bellstedt2020a}
Bellstedt S.,  et~al., 2020a, \mn@doi [MNRAS] {10.1093/mnras/staa1466}, 496,
  3235

\bibitem[\protect\citeauthoryear{Bellstedt et~al.,}{Bellstedt
  et~al.}{2020b}]{bellstedt2020b}
Bellstedt S.,  et~al., 2020b, \mn@doi [MNRAS] {10.1093/mnras/staa2620}, 498,
  5581

\bibitem[\protect\citeauthoryear{Bellstedt et~al.,}{Bellstedt
  et~al.}{2021}]{bellstedt2021}
Bellstedt S.,  et~al., 2021, \mn@doi [MNRAS] {10.1093/mnras/stab550}, 503, 3309

\bibitem[\protect\citeauthoryear{Bellstedt, Robotham, Driver, Lagos, Davies  \&
  Cook}{Bellstedt et~al.}{2024}]{bellstedt2024}
Bellstedt S.,  Robotham A. S.~G.,  Driver S.~P.,  Lagos C. d.~P.,  Davies L.
  J.~M.,   Cook R. H.~W.,  2024, \mn@doi [MNRAS] {10.1093/mnras/stae394}, 528,
  5452

\bibitem[\protect\citeauthoryear{Boquien, Burgarella, Roehlly, Buat, Ciesla,
  Corre, Inoue  \& Salas}{Boquien et~al.}{2019}]{boquien2019}
Boquien M.,  Burgarella D.,  Roehlly Y.,  Buat V.,  Ciesla L.,  Corre D.,
  Inoue A.~K.,   Salas H.,  2019, \mn@doi [A\&A] {10.1051/0004-6361/201834156},
  622, A103

\bibitem[\protect\citeauthoryear{Bouwens et~al.,}{Bouwens
  et~al.}{2012a}]{bouwens2012}
Bouwens R.~J.,  et~al., 2012a, \mn@doi [ApJL] {10.1088/2041-8205/752/1/L5},
  752, L5

\bibitem[\protect\citeauthoryear{Bouwens et~al.,}{Bouwens
  et~al.}{2012b}]{bouwens2012a}
Bouwens R.~J.,  et~al., 2012b, \mn@doi [ApJ] {10.1088/0004-637X/754/2/83}, 754,
  83

\bibitem[\protect\citeauthoryear{Bravo, Robotham, Lagos, Davies, Bellstedt  \&
  Thorne}{Bravo et~al.}{2022}]{bravo2022}
Bravo M.,  Robotham A. S.~G.,  Lagos C. d.~P.,  Davies L. J.~M.,  Bellstedt S.,
    Thorne J.~E.,  2022, \mn@doi [MNRAS] {10.1093/mnras/stac321}, 511, 5405

\bibitem[\protect\citeauthoryear{Bruzual \& Charlot}{Bruzual \&
  Charlot}{1993}]{bruzual1993}
Bruzual G.,  Charlot S.,  1993, \mn@doi [ApJ] {10.1086/172385}, 405, 538

\bibitem[\protect\citeauthoryear{Bruzual \& Charlot}{Bruzual \&
  Charlot}{2003}]{bruzual2003}
Bruzual G.,  Charlot S.,  2003, \mn@doi [MNRAS]
  {10.1046/j.1365-8711.2003.06897.x}, 344, 1000

\bibitem[\protect\citeauthoryear{Cappellari \& Emsellem}{Cappellari \&
  Emsellem}{2004}]{cappellari2004}
Cappellari M.,  Emsellem E.,  2004, \mn@doi [PASP] {10.1086/381875}, 116, 138

\bibitem[\protect\citeauthoryear{Carnall, McLure, Dunlop  \& Dav{\'e}}{Carnall
  et~al.}{2018}]{carnall2018a}
Carnall A.~C.,  McLure R.~J.,  Dunlop J.~S.,   Dav{\'e} R.,  2018, \mn@doi
  [MNRAS] {10.1093/mnras/sty2169}, 480, 4379

\bibitem[\protect\citeauthoryear{Carnall, Leja, Johnson, McLure, Dunlop  \&
  Conroy}{Carnall et~al.}{2019}]{carnall2019}
Carnall A.~C.,  Leja J.,  Johnson B.~D.,  McLure R.~J.,  Dunlop J.~S.,   Conroy
  C.,  2019, \mn@doi [ApJ] {10.3847/1538-4357/ab04a2}, 873, 44

\bibitem[\protect\citeauthoryear{Carnall et~al.,}{Carnall
  et~al.}{2023}]{carnall2023}
Carnall A.~C.,  et~al., 2023, \mn@doi [MNRAS] {10.1093/mnras/stad369}, 520,
  3974

\bibitem[\protect\citeauthoryear{Cassisi \& Salaris}{Cassisi \&
  Salaris}{1997}]{cassisi1997}
Cassisi S.,  Salaris M.,  1997, \mn@doi [MNRAS] {10.1093/mnras/285.3.593}, 285,
  593

\bibitem[\protect\citeauthoryear{Chabrier}{Chabrier}{2003}]{chabrier2003}
Chabrier G.,  2003, \mn@doi [PASP] {10.1086/376392}, 115, 763

\bibitem[\protect\citeauthoryear{Charlot \& Fall}{Charlot \&
  Fall}{2000}]{charlot2000}
Charlot S.,  Fall S.~M.,  2000, \mn@doi [ApJ] {10.1086/309250}, 539, 718

\bibitem[\protect\citeauthoryear{Charlot, Worthey  \& Bressan}{Charlot
  et~al.}{1996}]{charlot1996}
Charlot S.,  Worthey G.,   Bressan A.,  1996, \mn@doi [ApJ] {10.1086/176759},
  457, 625

\bibitem[\protect\citeauthoryear{Chevallard \& Charlot}{Chevallard \&
  Charlot}{2016}]{chevallard2016}
Chevallard J.,  Charlot S.,  2016, \mn@doi [MNRAS] {10.1093/mnras/stw1756},
  462, 1415

\bibitem[\protect\citeauthoryear{Cid~Fernandes, Mateus, Sodr{\'e},
  Stasi{\'n}ska  \& Gomes}{Cid~Fernandes et~al.}{2005}]{cidfernandes2005a}
Cid~Fernandes R.,  Mateus A.,  Sodr{\'e} L.,  Stasi{\'n}ska G.,   Gomes J.~M.,
  2005, \mn@doi [MNRAS] {10.1111/j.1365-2966.2005.08752.x}, 358, 363

\bibitem[\protect\citeauthoryear{Ciesla, Elbaz  \& Fensch}{Ciesla
  et~al.}{2017}]{ciesla2017}
Ciesla L.,  Elbaz D.,   Fensch J.,  2017, \mn@doi [A\&A]
  {10.1051/0004-6361/201731036}, 608, A41

\bibitem[\protect\citeauthoryear{Coelho, Bruzual  \& Charlot}{Coelho
  et~al.}{2020}]{coelho2020}
Coelho P. R.~T.,  Bruzual G.,   Charlot S.,  2020, \mn@doi [MNRAS]
  {10.1093/mnras/stz3023}, 491, 2025

\bibitem[\protect\citeauthoryear{Conroy}{Conroy}{2013}]{conroy2013}
Conroy C.,  2013, \mn@doi [ARA\&A] {10.1146/annurev-astro-082812-141017}, 51,
  393

\bibitem[\protect\citeauthoryear{Conroy \& Wechsler}{Conroy \&
  Wechsler}{2009}]{conroy2009}
Conroy C.,  Wechsler R.~H.,  2009, \mn@doi [ApJ] {10.1088/0004-637X/696/1/620},
  696, 620

\bibitem[\protect\citeauthoryear{Conroy, Gunn  \& White}{Conroy
  et~al.}{2009}]{conroy2009a}
Conroy C.,  Gunn J.~E.,   White M.,  2009, \mn@doi [ApJ]
  {10.1088/0004-637X/699/1/486}, 699, 486

\bibitem[\protect\citeauthoryear{Conroy, Villaume, {van Dokkum}  \&
  Lind}{Conroy et~al.}{2018}]{conroy2018}
Conroy C.,  Villaume A.,  {van Dokkum} P.~G.,   Lind K.,  2018, \mn@doi [ApJ]
  {10.3847/1538-4357/aaab49}, 854, 139

\bibitem[\protect\citeauthoryear{Cucciati et~al.,}{Cucciati
  et~al.}{2012}]{cucciati2012}
Cucciati O.,  et~al., 2012, \mn@doi [ApJ] {10.1051/0004-6361/201118010}, 539,
  A31

\bibitem[\protect\citeauthoryear{D'Silva et~al.,}{D'Silva
  et~al.}{2023}]{dsilva2023}
D'Silva J. C.~J.,  et~al., 2023, \mn@doi [MNRAS] {10.1093/mnras/stad1974}, 524,
  1448

\bibitem[\protect\citeauthoryear{Dahlen, Mobasher, Dickinson, Ferguson,
  Giavalisco, Kretchmer  \& Ravindranath}{Dahlen et~al.}{2007}]{dahlen2007}
Dahlen T.,  Mobasher B.,  Dickinson M.,  Ferguson H.~C.,  Giavalisco M.,
  Kretchmer C.,   Ravindranath S.,  2007, \mn@doi [ApJ] {10.1086/508854}, 654,
  172

\bibitem[\protect\citeauthoryear{Dale, Helou, Magdis, Armus, {D{\'i}az-Santos}
  \& Shi}{Dale et~al.}{2014}]{dale2014}
Dale D.~A.,  Helou G.,  Magdis G.~E.,  Armus L.,  {D{\'i}az-Santos} T.,   Shi
  Y.,  2014, \mn@doi [ApJ] {10.1088/0004-637X/784/1/83}, 784, 83

\bibitem[\protect\citeauthoryear{Dotter}{Dotter}{2016}]{dotter2016}
Dotter A.,  2016, \mn@doi [ApJSS] {10.3847/0067-0049/222/1/8}, 222, 8

\bibitem[\protect\citeauthoryear{Driver et~al.,}{Driver
  et~al.}{2011}]{driver2011}
Driver S.~P.,  et~al., 2011, \mn@doi [MNRAS]
  {10.1111/j.1365-2966.2010.18188.x}, 413, 971

\bibitem[\protect\citeauthoryear{Driver et~al.,}{Driver
  et~al.}{2018}]{driver2018}
Driver S.~P.,  et~al., 2018, \mn@doi [MNRAS] {10.1093/mnras/stx2728}, 475, 2891

\bibitem[\protect\citeauthoryear{Driver et~al.,}{Driver
  et~al.}{2022}]{driver2022}
Driver S.~P.,  et~al., 2022, \mn@doi [MNRAS] {10.1093/mnras/stac472}, 513, 439

\bibitem[\protect\citeauthoryear{Eldridge, Izzard  \& Tout}{Eldridge
  et~al.}{2008}]{eldridge2008}
Eldridge J.~J.,  Izzard R.~G.,   Tout C.~A.,  2008, \mn@doi [MNRAS]
  {10.1111/j.1365-2966.2007.12738.x}, 384, 1109

\bibitem[\protect\citeauthoryear{Girardi, Bertelli, Bressan, Chiosi,
  Groenewegen, Marigo, Salasnich  \& Weiss}{Girardi et~al.}{2002}]{girardi2002}
Girardi L.,  Bertelli G.,  Bressan A.,  Chiosi C.,  Groenewegen M. A.~T.,
  Marigo P.,  Salasnich B.,   Weiss A.,  2002, \mn@doi [A\&A]
  {10.1051/0004-6361:20020612}, 391, 195

\bibitem[\protect\citeauthoryear{Gruppioni et~al.,}{Gruppioni
  et~al.}{2013}]{gruppioni2013}
Gruppioni C.,  et~al., 2013, \mn@doi [MNRAS] {10.1093/mnras/stt308}, 432, 23

\bibitem[\protect\citeauthoryear{Han \& Han}{Han \& Han}{2012}]{han2012}
Han Y.,  Han Z.,  2012, \mn@doi [ApJ] {10.1088/0004-637X/749/2/123}, 749, 123

\bibitem[\protect\citeauthoryear{Harris et~al.,}{Harris
  et~al.}{2020}]{harris2020}
Harris C.~R.,  et~al., 2020, \mn@doi [Nature] {10.1038/s41586-020-2649-2}, 585,
  357

\bibitem[\protect\citeauthoryear{Haskell, Das, Smith, Cochrane, Hayward  \&
  {Angl{\'e}s-Alc{\'a}zar}}{Haskell et~al.}{2024}]{haskell2024}
Haskell P.,  Das S.,  Smith D. J.~B.,  Cochrane R.~K.,  Hayward C.~C.,
  {Angl{\'e}s-Alc{\'a}zar} D.,  2024, \mn@doi [MNRAS] {10.1093/mnrasl/slae019},
  530, L7

\bibitem[\protect\citeauthoryear{Hauschildt \& Baron}{Hauschildt \&
  Baron}{1999}]{hauschildt1999}
Hauschildt P.~H.,  Baron E.,  1999, \mn@doi [Journal of Computational and
  Applied Mathematics] {10.48550/arXiv.astro-ph/9808182}, 109, 41

\bibitem[\protect\citeauthoryear{Hidalgo et~al.,}{Hidalgo
  et~al.}{2018}]{hidalgo2018}
Hidalgo S.~L.,  et~al., 2018, \mn@doi [ApJ] {10.3847/1538-4357/aab158}, 856,
  125

\bibitem[\protect\citeauthoryear{Hunter}{Hunter}{2007}]{hunter2007}
Hunter J.~D.,  2007, \mn@doi [Computing in Science and Engineering]
  {10.1109/MCSE.2007.55}, 9, 90

\bibitem[\protect\citeauthoryear{Husser, {Wende-von Berg}, Dreizler, Homeier,
  Reiners, Barman  \& Hauschildt}{Husser et~al.}{2013}]{husser2013}
Husser T.~O.,  {Wende-von Berg} S.,  Dreizler S.,  Homeier D.,  Reiners A.,
  Barman T.,   Hauschildt P.~H.,  2013, \mn@doi [A\&A]
  {10.1051/0004-6361/201219058}, 553, A6

\bibitem[\protect\citeauthoryear{Iyer, Gawiser, Faber, Ferguson, Kartaltepe,
  Koekemoer, Pacifici  \& Somerville}{Iyer et~al.}{2019}]{iyer2019}
Iyer K.~G.,  Gawiser E.,  Faber S.~M.,  Ferguson H.~C.,  Kartaltepe J.,
  Koekemoer A.~M.,  Pacifici C.,   Somerville R.~S.,  2019, \mn@doi [ApJ]
  {10.3847/1538-4357/ab2052}, 879, 116

\bibitem[\protect\citeauthoryear{Iyer, Speagle, Caplar, Forbes, Gawiser, Leja
  \& Tacchella}{Iyer et~al.}{2024}]{iyer2024}
Iyer K.~G.,  Speagle J.~S.,  Caplar N.,  Forbes J.~C.,  Gawiser E.,  Leja J.,
  Tacchella S.,  2024, \mn@doi [ApJ] {10.3847/1538-4357/acff64}, 961, 53

\bibitem[\protect\citeauthoryear{Johnson, Leja, Conroy  \& Speagle}{Johnson
  et~al.}{2021}]{johnson2021}
Johnson B.~D.,  Leja J.,  Conroy C.,   Speagle J.~S.,  2021, \mn@doi [ApJSS]
  {10.3847/1538-4365/abef67}, 254, 22

\bibitem[\protect\citeauthoryear{Jones, Stanway  \& Carnall}{Jones
  et~al.}{2022}]{jones2022}
Jones G.~T.,  Stanway E.~R.,   Carnall A.~C.,  2022, \mn@doi [MNRAS]
  {10.1093/mnras/stac1667}, 514, 5706

\bibitem[\protect\citeauthoryear{Kennicutt}{Kennicutt}{1983}]{kennicutt1983}
Kennicutt Jr. R.~C.,  1983, \mn@doi [ApJ] {10.1086/161261}, 272, 54

\bibitem[\protect\citeauthoryear{Kroupa}{Kroupa}{2002}]{kroupa2002}
Kroupa P.,  2002, \mn@doi [Science] {10.1126/science.1067524}, 295, 82

\bibitem[\protect\citeauthoryear{Kurucz}{Kurucz}{1993}]{kurucz1993}
Kurucz R.~L.,  1993, {{SYNTHE}} Spectrum Synthesis Programs and Line Data

\bibitem[\protect\citeauthoryear{Lacey et~al.,}{Lacey et~al.}{2016}]{lacey2016}
Lacey C.~G.,  et~al., 2016, \mn@doi [MNRAS] {10.1093/mnras/stw1888}, 462, 3854

\bibitem[\protect\citeauthoryear{Lan{\c c}on \& Mouhcine}{Lan{\c c}on \&
  Mouhcine}{2002}]{lancon2002}
Lan{\c c}on A.,  Mouhcine M.,  2002, \mn@doi [A\&A]
  {10.1051/0004-6361:20020585}, 393, 167

\bibitem[\protect\citeauthoryear{Le~Borgne et~al.,}{Le~Borgne
  et~al.}{2003}]{leborgne2003}
Le~Borgne J.~F.,  et~al., 2003, \mn@doi [A\&A] {10.1051/0004-6361:20030243},
  402, 433

\bibitem[\protect\citeauthoryear{Leitherer et~al.,}{Leitherer
  et~al.}{1999}]{leitherer1999}
Leitherer C.,  et~al., 1999, \mn@doi [ApJSS] {10.1086/313233}, 123, 3

\bibitem[\protect\citeauthoryear{Leja, Johnson, Conroy, {van Dokkum}  \&
  Byler}{Leja et~al.}{2017}]{leja2017}
Leja J.,  Johnson B.~D.,  Conroy C.,  {van Dokkum} P.~G.,   Byler N.,  2017,
  \mn@doi [ApJ] {10.3847/1538-4357/aa5ffe}, 837, 170

\bibitem[\protect\citeauthoryear{Leja, Carnall, Johnson, Conroy  \&
  Speagle}{Leja et~al.}{2019}]{leja2019a}
Leja J.,  Carnall A.~C.,  Johnson B.~D.,  Conroy C.,   Speagle J.~S.,  2019,
  \mn@doi [ApJ] {10.3847/1538-4357/ab133c}, 876, 3

\bibitem[\protect\citeauthoryear{Lejeune \& Schaerer}{Lejeune \&
  Schaerer}{2001}]{lejeune2001}
Lejeune T.,  Schaerer D.,  2001, \mn@doi [A\&A] {10.1051/0004-6361:20000214},
  366, 538

\bibitem[\protect\citeauthoryear{Lejeune, Cuisinier  \& Buser}{Lejeune
  et~al.}{1997}]{lejeune1997}
Lejeune {\relax Th}.,  Cuisinier F.,   Buser R.,  1997, \mn@doi [A\&ASS]
  {10.1051/aas:1997373}, 125, 229

\bibitem[\protect\citeauthoryear{Liske et~al.,}{Liske et~al.}{2015}]{liske2015}
Liske J.,  et~al., 2015, \mn@doi [MNRAS] {10.1093/mnras/stv1436}, 452, 2087

\bibitem[\protect\citeauthoryear{Lower, Narayanan, Leja, Johnson, Conroy  \&
  Dav{\'e}}{Lower et~al.}{2020}]{lower2020}
Lower S.,  Narayanan D.,  Leja J.,  Johnson B.~D.,  Conroy C.,   Dav{\'e} R.,
  2020, \mn@doi [ApJ] {10.3847/1538-4357/abbfa7}, 904, 33

\bibitem[\protect\citeauthoryear{Madau \& Dickinson}{Madau \&
  Dickinson}{2014}]{madau2014}
Madau P.,  Dickinson M.,  2014, \mn@doi [ARA\&A]
  {10.1146/annurev-astro-081811-125615}, 52, 415

\bibitem[\protect\citeauthoryear{Magnelli, Elbaz, Chary, Dickinson, Le~Borgne,
  Frayer  \& Willmer}{Magnelli et~al.}{2011}]{magnelli2011}
Magnelli B.,  Elbaz D.,  Chary R.~R.,  Dickinson M.,  Le~Borgne D.,  Frayer
  D.~T.,   Willmer C. N.~A.,  2011, \mn@doi [ApJ]
  {10.1051/0004-6361/200913941}, 528, A35

\bibitem[\protect\citeauthoryear{Magnelli et~al.,}{Magnelli
  et~al.}{2013}]{magnelli2013}
Magnelli B.,  et~al., 2013, \mn@doi [ApJ] {10.1051/0004-6361/201321371}, 553,
  A132

\bibitem[\protect\citeauthoryear{Maraston}{Maraston}{1998}]{maraston1998}
Maraston C.,  1998, \mn@doi [MNRAS] {10.1046/j.1365-8711.1998.01947.x}, 300,
  872

\bibitem[\protect\citeauthoryear{Maraston}{Maraston}{2005}]{maraston2005}
Maraston C.,  2005, \mn@doi [MNRAS] {10.1111/j.1365-2966.2005.09270.x}, 362,
  799

\bibitem[\protect\citeauthoryear{Marigo \& Girardi}{Marigo \&
  Girardi}{2007}]{marigo2007}
Marigo P.,  Girardi L.,  2007, \mn@doi [A\&A] {10.1051/0004-6361:20066772},
  469, 239

\bibitem[\protect\citeauthoryear{Marigo, Girardi, Bressan, Groenewegen, Silva
  \& Granato}{Marigo et~al.}{2008}]{marigo2008}
Marigo P.,  Girardi L.,  Bressan A.,  Groenewegen M. A.~T.,  Silva L.,
  Granato G.~L.,  2008, \mn@doi [A\&A] {10.1051/0004-6361:20078467}, 482, 883

\bibitem[\protect\citeauthoryear{Marigo et~al.,}{Marigo
  et~al.}{2017}]{marigo2017}
Marigo P.,  et~al., 2017, \mn@doi [ApJ] {10.3847/1538-4357/835/1/77}, 835, 77

\bibitem[\protect\citeauthoryear{{Mart{\'i}n-Navarro}
  et~al.,}{{Mart{\'i}n-Navarro} et~al.}{2015}]{martin-navarro2015}
{Mart{\'i}n-Navarro} I.,  et~al., 2015, \mn@doi [ApJ]
  {10.1088/2041-8205/806/2/L31}, 806, L31

\bibitem[\protect\citeauthoryear{Martins, {Lima-Dias}, Coelho  \&
  Lagan{\'a}}{Martins et~al.}{2019}]{martins2019}
Martins L.~P.,  {Lima-Dias} C.,  Coelho P. R.~T.,   Lagan{\'a} T.~F.,  2019,
  \mn@doi [MNRAS] {10.1093/mnras/stz126}, 484, 2388

\bibitem[\protect\citeauthoryear{McKinney}{McKinney}{2010}]{mckinney2010}
McKinney W.,  2010, in van~der Walt S.,  Millman J.,  eds, Proceedings of the
  9th {{Python}} in {{Science Conference}}. pp 51--56

\bibitem[\protect\citeauthoryear{Mosleh, {Riahi-Zamin}  \& Tacchella}{Mosleh
  et~al.}{2025}]{mosleh2025}
Mosleh M.,  {Riahi-Zamin} M.,   Tacchella S.,  2025, Reconstructing {{Star
  Formation Histories}} of {{High-Redshift Galaxies}}: {{A Comparison}} of
  {{Resolved Parametric}} and {{Non-Parametric Models}} (\mn@eprint {arXiv}
  {2503.14591}), \mn@doi{10.48550/arXiv.2503.14591}

\bibitem[\protect\citeauthoryear{Nanayakkara et~al.,}{Nanayakkara
  et~al.}{2025}]{nanayakkara2025}
Nanayakkara T.,  et~al., 2025, \mn@doi [ApJ] {10.3847/1538-4357/ada6ac}, 981,
  78

\bibitem[\protect\citeauthoryear{Narayanan et~al.,}{Narayanan
  et~al.}{2025}]{narayanan2025}
Narayanan D.,  et~al., 2025, \mn@doi [ApJ] {10.3847/1538-4357/adb41c}, 982, 7

\bibitem[\protect\citeauthoryear{Osborne \& Salim}{Osborne \&
  Salim}{2024}]{osborne2024}
Osborne C.,  Salim S.,  2024, \mn@doi [ApJ] {10.3847/1538-4357/ad17c8}, 962, 59

\bibitem[\protect\citeauthoryear{Pacifici et~al.,}{Pacifici
  et~al.}{2023}]{pacifici2023}
Pacifici C.,  et~al., 2023, \mn@doi [ApJ] {10.3847/1538-4357/acacff}, 944, 141

\bibitem[\protect\citeauthoryear{Paspaliaris, Xilouris, Nersesian, Bianchi,
  Georgantopoulos, Masoura, Magdis  \& Plionis}{Paspaliaris
  et~al.}{2023}]{paspaliaris2023}
Paspaliaris E.-D.,  Xilouris E.~M.,  Nersesian A.,  Bianchi S.,
  Georgantopoulos I.,  Masoura V.~A.,  Magdis G.~E.,   Plionis M.,  2023,
  \mn@doi [A\&A] {10.1051/0004-6361/202244796}, 669, A11

\bibitem[\protect\citeauthoryear{Percival \& Salaris}{Percival \&
  Salaris}{2009}]{percival2009}
Percival S.~M.,  Salaris M.,  2009, \mn@doi [ApJ]
  {10.1088/0004-637X/703/1/1123}, 703, 1123

\bibitem[\protect\citeauthoryear{Pforr, Maraston  \& Tonini}{Pforr
  et~al.}{2012}]{pforr2012}
Pforr J.,  Maraston C.,   Tonini C.,  2012, \mn@doi [MNRAS]
  {10.1111/j.1365-2966.2012.20848.x}, 422, 3285

\bibitem[\protect\citeauthoryear{Pickles}{Pickles}{1998}]{pickles1998}
Pickles A.~J.,  1998, \mn@doi [PASP] {10.1086/316197}, 110, 863

\bibitem[\protect\citeauthoryear{Pietrinferni et~al.,}{Pietrinferni
  et~al.}{2021}]{pietrinferni2021}
Pietrinferni A.,  et~al., 2021, \mn@doi [ApJ] {10.3847/1538-4357/abd4d5}, 908,
  102

\bibitem[\protect\citeauthoryear{Pietrinferni, Salaris, Cassisi, Savino,
  Mucciarelli, Hyder  \& Hidalgo}{Pietrinferni et~al.}{2024}]{pietrinferni2024}
Pietrinferni A.,  Salaris M.,  Cassisi S.,  Savino A.,  Mucciarelli A.,  Hyder
  D.,   Hidalgo S.,  2024, \mn@doi [MNRAS] {10.1093/mnras/stad3267}, 527, 2065

\bibitem[\protect\citeauthoryear{{Planck Collaboration} et~al.,}{{Planck
  Collaboration} et~al.}{2016}]{planckcollaboration2016}
{Planck Collaboration} et~al., 2016, \mn@doi [A\&A]
  {10.1051/0004-6361/201525830}, 594, A13

\bibitem[\protect\citeauthoryear{Plat, Charlot, Bruzual, Feltre,
  {Vidal-Garc{\'i}a}, Morisset, Chevallard  \& Todt}{Plat
  et~al.}{2019}]{plat2019}
Plat A.,  Charlot S.,  Bruzual G.,  Feltre A.,  {Vidal-Garc{\'i}a} A.,
  Morisset C.,  Chevallard J.,   Todt H.,  2019, \mn@doi [MNRAS]
  {10.1093/mnras/stz2616}, 490, 978

\bibitem[\protect\citeauthoryear{{R Core Team}}{{R Core
  Team}}{2017}]{rcoreteam2017}
{R Core Team} 2017, R: {{A Language}} and {{Environment}} for {{Statistical
  Computing}}.
R Foundation for Statistical Computing, Vienna, Austria

\bibitem[\protect\citeauthoryear{Reddy \& Steidel}{Reddy \&
  Steidel}{2009}]{reddy2009}
Reddy N.~A.,  Steidel C.~C.,  2009, \mn@doi [ApJ]
  {10.1088/0004-637X/692/1/778}, 692, 778

\bibitem[\protect\citeauthoryear{Robotham \& Bellstedt}{Robotham \&
  Bellstedt}{2024}]{robotham2024a}
Robotham A. S.~G.,  Bellstedt S.,  2024, {{ProGeny I}}: A New Simple Stellar
  Population Generator and Impact of Isochrones / Stellar Atmospheres / Initial
  Mass Functions, \mn@doi{10.48550/arXiv.2410.17697}

\bibitem[\protect\citeauthoryear{Robotham \& Driver}{Robotham \&
  Driver}{2011}]{robotham2011}
Robotham A. S.~G.,  Driver S.~P.,  2011, \mn@doi [MNRAS]
  {10.1111/j.1365-2966.2011.18327.x}, 413, 2570

\bibitem[\protect\citeauthoryear{Robotham, Bellstedt, Lagos, Thorne, Davies,
  Driver  \& Bravo}{Robotham et~al.}{2020}]{robotham2020}
Robotham A. S.~G.,  Bellstedt S.,  Lagos C. d.~P.,  Thorne J.~E.,  Davies
  L.~J.,  Driver S.~P.,   Bravo M.,  2020, \mn@doi [MNRAS]
  {10.1093/mnras/staa1116}, 495, 905

\bibitem[\protect\citeauthoryear{Salpeter}{Salpeter}{1955}]{salpeter1955}
Salpeter E.~E.,  1955, \mn@doi [ApJ] {10.1086/145971}, 121, 161

\bibitem[\protect\citeauthoryear{{S{\'a}nchez-Bl{\'a}zquez}
  et~al.,}{{S{\'a}nchez-Bl{\'a}zquez} et~al.}{2006}]{sanchez-blazquez2006}
{S{\'a}nchez-Bl{\'a}zquez} P.,  et~al., 2006, \mn@doi [MNRAS]
  {10.1111/j.1365-2966.2006.10699.x}, 371, 703

\bibitem[\protect\citeauthoryear{Sanders}{Sanders}{2003}]{sanders2003}
Sanders R.~H.,  2003, \mn@doi [MNRAS] {10.1046/j.1365-8711.2003.06596.x}, 342,
  901

\bibitem[\protect\citeauthoryear{Schaller, Schaerer, Meynet  \&
  Maeder}{Schaller et~al.}{1992}]{schaller1992}
Schaller G.,  Schaerer D.,  Meynet G.,   Maeder A.,  1992, MNRAS, 96, 269

\bibitem[\protect\citeauthoryear{Schenker et~al.,}{Schenker
  et~al.}{2013}]{schenker2013}
Schenker M.~A.,  et~al., 2013, \mn@doi [ApJ] {10.1088/0004-637X/768/2/196},
  768, 196

\bibitem[\protect\citeauthoryear{Schiminovich et~al.,}{Schiminovich
  et~al.}{2005}]{schiminovich2005}
Schiminovich D.,  et~al., 2005, \mn@doi [ApJL] {10.1086/427077}, 619, L47

\bibitem[\protect\citeauthoryear{Shivaei et~al.,}{Shivaei
  et~al.}{2024}]{shivaei2024}
Shivaei I.,  et~al., 2024, \mn@doi [A\&A] {10.1051/0004-6361/202449579}, 690,
  A89

\bibitem[\protect\citeauthoryear{Stanway \& Eldridge}{Stanway \&
  Eldridge}{2018}]{stanway2018}
Stanway E.~R.,  Eldridge J.~J.,  2018, \mn@doi [MNRAS] {10.1093/mnras/sty1353},
  479, 75

\bibitem[\protect\citeauthoryear{Suess et~al.,}{Suess et~al.}{2022}]{suess2022}
Suess K.~A.,  et~al., 2022, \mn@doi [ApJ] {10.3847/1538-4357/ac82b0}, 935, 146

\bibitem[\protect\citeauthoryear{Sun, {Faucher-Gigu{\`e}re}, Hayward, Shen,
  Wetzel  \& Cochrane}{Sun et~al.}{2023}]{sun2023}
Sun G.,  {Faucher-Gigu{\`e}re} C.-A.,  Hayward C.~C.,  Shen X.,  Wetzel A.,
  Cochrane R.~K.,  2023, \mn@doi [ApJ] {10.3847/2041-8213/acf85a}, 955, L35

\bibitem[\protect\citeauthoryear{Takeuchi, Yoshikawa  \& Ishii}{Takeuchi
  et~al.}{2003}]{takeuchi2003}
Takeuchi T.~T.,  Yoshikawa K.,   Ishii T.~T.,  2003, \mn@doi [ApJL]
  {10.1086/375181}, 587, L89

\bibitem[\protect\citeauthoryear{Thomas, Maraston  \& Bender}{Thomas
  et~al.}{2003}]{thomas2003}
Thomas D.,  Maraston C.,   Bender R.,  2003, \mn@doi [MNRAS]
  {10.1046/j.1365-8711.2003.06248.x}, 339, 897

\bibitem[\protect\citeauthoryear{Thorne et~al.,}{Thorne
  et~al.}{2021}]{thorne2021}
Thorne J.~E.,  et~al., 2021, \mn@doi [MNRAS] {10.1093/mnras/stab1294}, 505, 540

\bibitem[\protect\citeauthoryear{Thorne et~al.,}{Thorne
  et~al.}{2022a}]{thorne2022}
Thorne J.~E.,  et~al., 2022a, \mn@doi [MNRAS] {10.1093/mnras/stab3208}, 509,
  4940

\bibitem[\protect\citeauthoryear{Thorne et~al.,}{Thorne
  et~al.}{2022b}]{thorne2022a}
Thorne J.~E.,  et~al., 2022b, \mn@doi [MNRAS] {10.1093/mnras/stac3082}, 517,
  6035

\bibitem[\protect\citeauthoryear{Thorne, Robotham, Bellstedt  \& Davies}{Thorne
  et~al.}{2023}]{thorne2023}
Thorne J.~E.,  Robotham A. S.~G.,  Bellstedt S.,   Davies L. J.~M.,  2023,
  \mn@doi [MNRAS] {10.1093/mnras/stad1361}, 522, 6354

\bibitem[\protect\citeauthoryear{Tinsley}{Tinsley}{1978}]{tinsley1978}
Tinsley B.~M.,  1978, \mn@doi [ApJ] {10.1086/156116}, 222, 14

\bibitem[\protect\citeauthoryear{Tinsley \& Gunn}{Tinsley \&
  Gunn}{1976}]{tinsley1976}
Tinsley B.~M.,  Gunn J.~E.,  1976, \mn@doi [ApJ] {10.1086/154046}, 203, 52

\bibitem[\protect\citeauthoryear{Tremonti et~al.,}{Tremonti
  et~al.}{2004}]{tremonti2004}
Tremonti C.~A.,  et~al., 2004, \mn@doi [ApJ] {10.1086/423264}, 613, 898

\bibitem[\protect\citeauthoryear{Vazdekis, Koleva, Ricciardelli, R{\"o}ck  \&
  {Falc{\'o}n-Barroso}}{Vazdekis et~al.}{2016}]{vazdekis2016}
Vazdekis A.,  Koleva M.,  Ricciardelli E.,  R{\"o}ck B.,   {Falc{\'o}n-Barroso}
  J.,  2016, \mn@doi [MNRAS] {10.1093/mnras/stw2231}, 463, 3409

\bibitem[\protect\citeauthoryear{Verro et~al.,}{Verro et~al.}{2022}]{verro2022}
Verro K.,  et~al., 2022, \mn@doi [A\&A] {10.1051/0004-6361/202142387}, 661, A50

\bibitem[\protect\citeauthoryear{Walcher, Groves, Budav{\'a}ri  \&
  Dale}{Walcher et~al.}{2011}]{walcher2011}
Walcher J.,  Groves B.,  Budav{\'a}ri T.,   Dale D.,  2011, \mn@doi
  [Astrophysics and Space Science] {10.1007/s10509-010-0458-z}, 331, 1

\bibitem[\protect\citeauthoryear{Wilkinson, Maraston, Goddard, Thomas  \&
  Parikh}{Wilkinson et~al.}{2017}]{wilkinson2017}
Wilkinson D.~M.,  Maraston C.,  Goddard D.,  Thomas D.,   Parikh T.,  2017,
  \mn@doi [MNRAS] {10.1093/mnras/stx2215}, 472, 4297

\bibitem[\protect\citeauthoryear{Wyder et~al.,}{Wyder et~al.}{2005}]{wyder2005}
Wyder T.~K.,  et~al., 2005, \mn@doi [ApJL] {10.1086/424735}, 619, L15

\bibitem[\protect\citeauthoryear{Yang et~al.,}{Yang et~al.}{2022}]{yang2022}
Yang G.,  et~al., 2022, \mn@doi [ApJ] {10.3847/1538-4357/ac4971}, 927, 192

\bibitem[\protect\citeauthoryear{{da Cunha}, Charlot  \& Elbaz}{{da Cunha}
  et~al.}{2008}]{dacunha2008}
{da Cunha} E.,  Charlot S.,   Elbaz D.,  2008, \mn@doi [MNRAS]
  {10.1111/j.1365-2966.2008.13535.x}, 388, 1595

\bibitem[\protect\citeauthoryear{{van der Velden}}{{van der
  Velden}}{2020}]{vandervelden2020}
{van der Velden} E.,  2020, \mn@doi [The Journal of Open Source Software]
  {10.21105/joss.02004}, 5, 2004

\makeatother
\end{thebibliography}


\end{document}